\hsize=31pc 
\vsize=49pc 
\lineskip=0pt 
\parskip=0pt plus 1pt 
\hfuzz=1pt   
\vfuzz=2pt 
\pretolerance=2500 
\tolerance=5000 
\vbadness=5000 
\hbadness=5000 
\widowpenalty=500 
\clubpenalty=200 
\brokenpenalty=500 
\predisplaypenalty=200 
\voffset=-1pc 
\nopagenumbers      
\catcode`@=11 
\newif\ifams 
\amsfalse 
%
%
%
\newfam\bdifam 
\newfam\bsyfam 
\newfam\bssfam 
\newfam\msafam 
\newfam\msbfam 
\newif\ifxxpt    
\newif\ifxviipt  
\newif\ifxivpt   
\newif\ifxiipt   
\newif\ifxipt    
\newif\ifxpt     
\newif\ifixpt    
\newif\ifviiipt  
\newif\ifviipt   
\newif\ifvipt    
\newif\ifvpt     
%
%
\def\headsize#1#2{\def\headb@seline{#2}%
                \ifnum#1=20\def\HEAD{twenty}%
                           \def\smHEAD{twelve}%
                           \def\vsHEAD{nine}%
                           \ifxxpt\else\xdef\f@ntsize{\HEAD}%
                           \def\m@g{4}\def\s@ze{20.74}%
                           \loadheadfonts\xxpttrue\fi 
                           \ifxiipt\else\xdef\f@ntsize{\smHEAD}%
                           \def\m@g{1}\def\s@ze{12}%
                           \loadxiiptfonts\xiipttrue\fi 
                           \ifixpt\else\xdef\f@ntsize{\vsHEAD}%
                           \def\s@ze{9}%
                           \loadsmallfonts\ixpttrue\fi 
                      \else 
                \ifnum#1=17\def\HEAD{seventeen}%
                           \def\smHEAD{eleven}%
                           \def\vsHEAD{eight}%
                           \ifxviipt\else\xdef\f@ntsize{\HEAD}%
                           \def\m@g{3}\def\s@ze{17.28}%
                           \loadheadfonts\xviipttrue\fi 
                           \ifxipt\else\xdef\f@ntsize{\smHEAD}%
                           \loadxiptfonts\xipttrue\fi 
                           \ifviiipt\else\xdef\f@ntsize{\vsHEAD}%
                           \def\s@ze{8}%
                           \loadsmallfonts\viiipttrue\fi 
                      \else\def\HEAD{fourteen}%
                           \def\smHEAD{ten}%
                           \def\vsHEAD{seven}%
                           \ifxivpt\else\xdef\f@ntsize{\HEAD}%
                           \def\m@g{2}\def\s@ze{14.4}%
                           \loadheadfonts\xivpttrue\fi 
                           \ifxpt\else\xdef\f@ntsize{\smHEAD}%
                           \def\s@ze{10}%
                           \loadxptfonts\xpttrue\fi 
                           \ifviipt\else\xdef\f@ntsize{\vsHEAD}%
                           \def\s@ze{7}%
                           \loadviiptfonts\viipttrue\fi 
                \ifnum#1=14\else 
                \message{Header size should be 20, 17 or 14 point 
                              will now default to 14pt}\fi 
                \fi\fi\headfonts} 
%
%
\def\textsize#1#2{\def\textb@seline{#2}%
                 \ifnum#1=12\def\TEXT{twelve}%
                           \def\smTEXT{eight}%
                           \def\vsTEXT{six}%
                           \ifxiipt\else\xdef\f@ntsize{\TEXT}%
                           \def\m@g{1}\def\s@ze{12}%
                           \loadxiiptfonts\xiipttrue\fi 
                           \ifviiipt\else\xdef\f@ntsize{\smTEXT}%
                           \def\s@ze{8}%
                           \loadsmallfonts\viiipttrue\fi 
                           \ifvipt\else\xdef\f@ntsize{\vsTEXT}%
                           \def\s@ze{6}%
                           \loadviptfonts\vipttrue\fi 
                      \else 
                \ifnum#1=11\def\TEXT{eleven}%
                           \def\smTEXT{seven}%
                           \def\vsTEXT{five}%
                           \ifxipt\else\xdef\f@ntsize{\TEXT}%
                           \def\s@ze{11}%
                           \loadxiptfonts\xipttrue\fi 
                           \ifviipt\else\xdef\f@ntsize{\smTEXT}%
                           \loadviiptfonts\viipttrue\fi 
                           \ifvpt\else\xdef\f@ntsize{\vsTEXT}%
                           \def\s@ze{5}%
                           \loadvptfonts\vpttrue\fi 
                      \else\def\TEXT{ten}%
                           \def\smTEXT{seven}%
                           \def\vsTEXT{five}%
                           \ifxpt\else\xdef\f@ntsize{\TEXT}%
                           \loadxptfonts\xpttrue\fi 
                           \ifviipt\else\xdef\f@ntsize{\smTEXT}%
                           \def\s@ze{7}%
                           \loadviiptfonts\viipttrue\fi 
                           \ifvpt\else\xdef\f@ntsize{\vsTEXT}%
                           \def\s@ze{5}%
                           \loadvptfonts\vpttrue\fi 
                \ifnum#1=10\else 
                \message{Text size should be 12, 11 or 10 point 
                              will now default to 10pt}\fi 
                \fi\fi\textfonts} 
%
%
\def\smallsize#1#2{\def\smallb@seline{#2}%
                 \ifnum#1=10\def\SMALL{ten}%
                           \def\smSMALL{seven}%
                           \def\vsSMALL{five}%
                           \ifxpt\else\xdef\f@ntsize{\SMALL}%
                           \loadxptfonts\xpttrue\fi 
                           \ifviipt\else\xdef\f@ntsize{\smSMALL}%
                           \def\s@ze{7}%
                           \loadviiptfonts\viipttrue\fi 
                           \ifvpt\else\xdef\f@ntsize{\vsSMALL}%
                           \def\s@ze{5}%
                           \loadvptfonts\vpttrue\fi 
                       \else 
                 \ifnum#1=9\def\SMALL{nine}%
                           \def\smSMALL{six}%
                           \def\vsSMALL{five}%
                           \ifixpt\else\xdef\f@ntsize{\SMALL}%
                           \def\s@ze{9}%
                           \loadsmallfonts\ixpttrue\fi 
                           \ifvipt\else\xdef\f@ntsize{\smSMALL}%
                           \def\s@ze{6}%
                           \loadviptfonts\vipttrue\fi 
                           \ifvpt\else\xdef\f@ntsize{\vsSMALL}%
                           \def\s@ze{5}%
                           \loadvptfonts\vpttrue\fi 
                       \else 
                           \def\SMALL{eight}%
                           \def\smSMALL{six}%
                           \def\vsSMALL{five}%
                           \ifviiipt\else\xdef\f@ntsize{\SMALL}%
                           \def\s@ze{8}%
                           \loadsmallfonts\viiipttrue\fi 
                           \ifvipt\else\xdef\f@ntsize{\smSMALL}%
                           \def\s@ze{6}%
                           \loadviptfonts\vipttrue\fi 
                           \ifvpt\else\xdef\f@ntsize{\vsSMALL}%
                           \def\s@ze{5}%
                           \loadvptfonts\vpttrue\fi 
                 \ifnum#1=8\else\message{Small size should be 10, 9 or  
                            8 point will now default to 8pt}\fi 
                \fi\fi\smallfonts} 
\def\F@nt{\expandafter\font\csname} 
\def\Sk@w{\expandafter\skewchar\csname} 
\def\@nd{\endcsname} 
\def\@step#1{ scaled \magstep#1} 
\def\@half{ scaled \magstephalf} 
\def\@t#1{ at #1pt} 
%
%
\def\loadheadfonts{\bigf@nts 
\F@nt \f@ntsize bdi\@nd=cmmib10 \@t{\s@ze}%
\Sk@w \f@ntsize bdi\@nd='177 
\F@nt \f@ntsize bsy\@nd=cmbsy10 \@t{\s@ze}%
\Sk@w \f@ntsize bsy\@nd='60 
\F@nt \f@ntsize bss\@nd=cmssbx10 \@t{\s@ze}} 
%
%
\def\loadxiiptfonts{\bigf@nts 
\F@nt \f@ntsize bdi\@nd=cmmib10 \@step{\m@g}%
\Sk@w \f@ntsize bdi\@nd='177 
\F@nt \f@ntsize bsy\@nd=cmbsy10 \@step{\m@g}%
\Sk@w \f@ntsize bsy\@nd='60 
\F@nt \f@ntsize bss\@nd=cmssbx10 \@step{\m@g}} 
%
%
\def\loadxiptfonts{%
\font\elevenrm=cmr10 \@half 
\font\eleveni=cmmi10 \@half 
\skewchar\eleveni='177 
\font\elevensy=cmsy10 \@half 
\skewchar\elevensy='60 
\font\elevenex=cmex10 \@half 
\font\elevenit=cmti10 \@half 
\font\elevensl=cmsl10 \@half 
\font\elevenbf=cmbx10 \@half 
\font\eleventt=cmtt10 \@half 
\ifams\font\elevenmsa=msam10 \@half 
\font\elevenmsb=msbm10 \@half\else\fi 
\font\elevenbdi=cmmib10 \@half 
\skewchar\elevenbdi='177 
\font\elevenbsy=cmbsy10 \@half 
\skewchar\elevenbsy='60 
\font\elevenbss=cmssbx10 \@half} 
%
%
\def\loadxptfonts{%
\font\tenbdi=cmmib10 
\skewchar\tenbdi='177 
\font\tenbsy=cmbsy10  
\skewchar\tenbsy='60 
\ifams\font\tenmsa=msam10  
\font\tenmsb=msbm10\else\fi 
\font\tenbss=cmssbx10}%
%
%
\def\loadsmallfonts{\smallf@nts 
\ifams 
\F@nt \f@ntsize ex\@nd=cmex\s@ze 
\else 
\F@nt \f@ntsize ex\@nd=cmex10\fi 
\F@nt \f@ntsize it\@nd=cmti\s@ze 
\F@nt \f@ntsize sl\@nd=cmsl\s@ze 
\F@nt \f@ntsize tt\@nd=cmtt\s@ze} 
%
%
\def\loadviiptfonts{%
\font\sevenit=cmti7 
\font\sevensl=cmsl8 at 7pt 
\ifams\font\sevenmsa=msam7  
\font\sevenmsb=msbm7 
\font\sevenex=cmex7 
\font\sevenbsy=cmbsy7 
\font\sevenbdi=cmmib7\else 
\font\sevenex=cmex10 
\font\sevenbsy=cmbsy10 at 7pt 
\font\sevenbdi=cmmib10 at 7pt\fi 
\skewchar\sevenbsy='60 
\skewchar\sevenbdi='177 
\font\sevenbss=cmssbx10 at 7pt}%
%
%
\def\loadviptfonts{\smallf@nts 
\ifams\font\sixex=cmex7 at 6pt\else 
\font\sixex=cmex10\fi 
\font\sixit=cmti7 at 6pt} 
%
%
\def\loadvptfonts{%
\font\fiveit=cmti7 at 5pt 
\ifams\font\fiveex=cmex7 at 5pt 
\font\fivebdi=cmmib5 
\font\fivebsy=cmbsy5 
\font\fivemsa=msam5  
\font\fivemsb=msbm5\else 
\font\fiveex=cmex10 
\font\fivebdi=cmmib10 at 5pt 
\font\fivebsy=cmbsy10 at 5pt\fi 
\skewchar\fivebdi='177 
\skewchar\fivebsy='60 
\font\fivebss=cmssbx10 at 5pt} 
\def\bigf@nts{%
\F@nt \f@ntsize rm\@nd=cmr10 \@step{\m@g}%
\F@nt \f@ntsize i\@nd=cmmi10 \@step{\m@g}%
\Sk@w \f@ntsize i\@nd='177 
\F@nt \f@ntsize sy\@nd=cmsy10 \@step{\m@g}%
\Sk@w \f@ntsize sy\@nd='60 
\F@nt \f@ntsize ex\@nd=cmex10 \@step{\m@g}%
\F@nt \f@ntsize it\@nd=cmti10 \@step{\m@g}%
\F@nt \f@ntsize sl\@nd=cmsl10 \@step{\m@g}%
\F@nt \f@ntsize bf\@nd=cmbx10 \@step{\m@g}%
\F@nt \f@ntsize tt\@nd=cmtt10 \@step{\m@g}%
\ifams 
\F@nt \f@ntsize msa\@nd=msam10 \@step{\m@g}%
\F@nt \f@ntsize msb\@nd=msbm10 \@step{\m@g}\else\fi} 
\def\smallf@nts{%
\F@nt \f@ntsize rm\@nd=cmr\s@ze 
\F@nt \f@ntsize i\@nd=cmmi\s@ze  
\Sk@w \f@ntsize i\@nd='177 
\F@nt \f@ntsize sy\@nd=cmsy\s@ze 
\Sk@w \f@ntsize sy\@nd='60 
\F@nt \f@ntsize bf\@nd=cmbx\s@ze  
\ifams 
\F@nt \f@ntsize bdi\@nd=cmmib\s@ze  
\F@nt \f@ntsize bsy\@nd=cmbsy\s@ze  
\F@nt \f@ntsize msa\@nd=msam\s@ze  
\F@nt \f@ntsize msb\@nd=msbm\s@ze 
\else 
\F@nt \f@ntsize bdi\@nd=cmmib10 \@t{\s@ze}%
\F@nt \f@ntsize bsy\@nd=cmbsy10 \@t{\s@ze}\fi  
\Sk@w \f@ntsize bdi\@nd='177 
\Sk@w \f@ntsize bsy\@nd='60 
\F@nt \f@ntsize bss\@nd=cmssbx10 \@t{\s@ze}}%
%
%
\def\headfonts{%
\textfont0=\csname\HEAD rm\@nd         
\scriptfont0=\csname\smHEAD rm\@nd 
\scriptscriptfont0=\csname\vsHEAD rm\@nd 
\def\rm{\fam0\csname\HEAD rm\@nd 
\def\sc{\csname\smHEAD rm\@nd}}%
\textfont1=\csname\HEAD i\@nd          
\scriptfont1=\csname\smHEAD i\@nd 
\scriptscriptfont1=\csname\vsHEAD i\@nd 
\textfont2=\csname\HEAD sy\@nd         
\scriptfont2=\csname\smHEAD sy\@nd 
\scriptscriptfont2=\csname\vsHEAD sy\@nd 
\textfont3=\csname\HEAD ex\@nd         
\scriptfont3=\csname\smHEAD ex\@nd 
\scriptscriptfont3=\csname\smHEAD ex\@nd 
\textfont\itfam=\csname\HEAD it\@nd    
\scriptfont\itfam=\csname\smHEAD it\@nd 
\scriptscriptfont\itfam=\csname\vsHEAD it\@nd 
\def\it{\fam\itfam\csname\HEAD it\@nd 
\def\sc{\csname\smHEAD it\@nd}}%
\textfont\slfam=\csname\HEAD sl\@nd    
\def\sl{\fam\slfam\csname\HEAD sl\@nd 
\def\sc{\csname\smHEAD sl\@nd}}%
\textfont\bffam=\csname\HEAD bf\@nd    
\scriptfont\bffam=\csname\smHEAD bf\@nd 
\scriptscriptfont\bffam=\csname\vsHEAD bf\@nd 
\def\bf{\fam\bffam\csname\HEAD bf\@nd 
\def\sc{\csname\smHEAD bf\@nd}}%
\textfont\ttfam=\csname\HEAD tt\@nd    
\def\tt{\fam\ttfam\csname\HEAD tt\@nd}%
\textfont\bdifam=\csname\HEAD bdi\@nd  
\scriptfont\bdifam=\csname\smHEAD bdi\@nd 
\scriptscriptfont\bdifam=\csname\vsHEAD bdi\@nd 
\def\bdi{\fam\bdifam\csname\HEAD bdi\@nd}%
\textfont\bsyfam=\csname\HEAD bsy\@nd  
\scriptfont\bsyfam=\csname\smHEAD bsy\@nd 
\def\bsy{\fam\bsyfam\csname\HEAD bsy\@nd}%
\textfont\bssfam=\csname\HEAD bss\@nd  
\scriptfont\bssfam=\csname\smHEAD bss\@nd 
\scriptscriptfont\bssfam=\csname\vsHEAD bss\@nd 
\def\bss{\fam\bssfam\csname\HEAD bss\@nd}%
\ifams 
\textfont\msafam=\csname\HEAD msa\@nd  
\scriptfont\msafam=\csname\smHEAD msa\@nd 
\scriptscriptfont\msafam=\csname\vsHEAD msa\@nd 
\textfont\msbfam=\csname\HEAD msb\@nd  
\scriptfont\msbfam=\csname\smHEAD msb\@nd 
\scriptscriptfont\msbfam=\csname\vsHEAD msb\@nd 
\else\fi 
\normalbaselineskip=\headb@seline pt%
\setbox\strutbox=\hbox{\vrule height.7\normalbaselineskip  
depth.3\baselineskip width0pt}%
\def\sc{\csname\smHEAD rm\@nd}\normalbaselines\bf} 
%
%
\def\textfonts{%
\textfont0=\csname\TEXT rm\@nd         
\scriptfont0=\csname\smTEXT rm\@nd 
\scriptscriptfont0=\csname\vsTEXT rm\@nd 
\def\rm{\fam0\csname\TEXT rm\@nd 
\def\sc{\csname\smTEXT rm\@nd}}%
\textfont1=\csname\TEXT i\@nd          
\scriptfont1=\csname\smTEXT i\@nd 
\scriptscriptfont1=\csname\vsTEXT i\@nd 
\textfont2=\csname\TEXT sy\@nd         
\scriptfont2=\csname\smTEXT sy\@nd 
\scriptscriptfont2=\csname\vsTEXT sy\@nd 
\textfont3=\csname\TEXT ex\@nd         
\scriptfont3=\csname\smTEXT ex\@nd 
\scriptscriptfont3=\csname\smTEXT ex\@nd 
\textfont\itfam=\csname\TEXT it\@nd    
\scriptfont\itfam=\csname\smTEXT it\@nd 
\scriptscriptfont\itfam=\csname\vsTEXT it\@nd 
\def\it{\fam\itfam\csname\TEXT it\@nd 
\def\sc{\csname\smTEXT it\@nd}}%
\textfont\slfam=\csname\TEXT sl\@nd    
\def\sl{\fam\slfam\csname\TEXT sl\@nd 
\def\sc{\csname\smTEXT sl\@nd}}%
\textfont\bffam=\csname\TEXT bf\@nd    
\scriptfont\bffam=\csname\smTEXT bf\@nd 
\scriptscriptfont\bffam=\csname\vsTEXT bf\@nd 
\def\bf{\fam\bffam\csname\TEXT bf\@nd 
\def\sc{\csname\smTEXT bf\@nd}}%
\textfont\ttfam=\csname\TEXT tt\@nd    
\def\tt{\fam\ttfam\csname\TEXT tt\@nd}%
\textfont\bdifam=\csname\TEXT bdi\@nd  
\scriptfont\bdifam=\csname\smTEXT bdi\@nd 
\scriptscriptfont\bdifam=\csname\vsTEXT bdi\@nd 
\def\bdi{\fam\bdifam\csname\TEXT bdi\@nd}%
\textfont\bsyfam=\csname\TEXT bsy\@nd  
\scriptfont\bsyfam=\csname\smTEXT bsy\@nd 
\def\bsy{\fam\bsyfam\csname\TEXT bsy\@nd}%
\textfont\bssfam=\csname\TEXT bss\@nd  
\scriptfont\bssfam=\csname\smTEXT bss\@nd 
\scriptscriptfont\bssfam=\csname\vsTEXT bss\@nd 
\def\bss{\fam\bssfam\csname\TEXT bss\@nd}%
\ifams 
\textfont\msafam=\csname\TEXT msa\@nd  
\scriptfont\msafam=\csname\smTEXT msa\@nd 
\scriptscriptfont\msafam=\csname\vsTEXT msa\@nd 
\textfont\msbfam=\csname\TEXT msb\@nd  
\scriptfont\msbfam=\csname\smTEXT msb\@nd 
\scriptscriptfont\msbfam=\csname\vsTEXT msb\@nd 
\else\fi 
\normalbaselineskip=\textb@seline pt 
\setbox\strutbox=\hbox{\vrule height.7\normalbaselineskip  
depth.3\baselineskip width0pt}%
\everymath{}%
\def\sc{\csname\smTEXT rm\@nd}\normalbaselines\rm} 
%
%
\def\smallfonts{%
\textfont0=\csname\SMALL rm\@nd         
\scriptfont0=\csname\smSMALL rm\@nd 
\scriptscriptfont0=\csname\vsSMALL rm\@nd 
\def\rm{\fam0\csname\SMALL rm\@nd 
\def\sc{\csname\smSMALL rm\@nd}}%
\textfont1=\csname\SMALL i\@nd          
\scriptfont1=\csname\smSMALL i\@nd 
\scriptscriptfont1=\csname\vsSMALL i\@nd 
\textfont2=\csname\SMALL sy\@nd         
\scriptfont2=\csname\smSMALL sy\@nd 
\scriptscriptfont2=\csname\vsSMALL sy\@nd 
\textfont3=\csname\SMALL ex\@nd         
\scriptfont3=\csname\smSMALL ex\@nd 
\scriptscriptfont3=\csname\smSMALL ex\@nd 
\textfont\itfam=\csname\SMALL it\@nd    
\scriptfont\itfam=\csname\smSMALL it\@nd 
\scriptscriptfont\itfam=\csname\vsSMALL it\@nd 
\def\it{\fam\itfam\csname\SMALL it\@nd 
\def\sc{\csname\smSMALL it\@nd}}%
\textfont\slfam=\csname\SMALL sl\@nd    
\def\sl{\fam\slfam\csname\SMALL sl\@nd 
\def\sc{\csname\smSMALL sl\@nd}}%
\textfont\bffam=\csname\SMALL bf\@nd    
\scriptfont\bffam=\csname\smSMALL bf\@nd 
\scriptscriptfont\bffam=\csname\vsSMALL bf\@nd 
\def\bf{\fam\bffam\csname\SMALL bf\@nd 
\def\sc{\csname\smSMALL bf\@nd}}%
\textfont\ttfam=\csname\SMALL tt\@nd    
\def\tt{\fam\ttfam\csname\SMALL tt\@nd}%
\textfont\bdifam=\csname\SMALL bdi\@nd  
\scriptfont\bdifam=\csname\smSMALL bdi\@nd 
\scriptscriptfont\bdifam=\csname\vsSMALL bdi\@nd 
\def\bdi{\fam\bdifam\csname\SMALL bdi\@nd}%
\textfont\bsyfam=\csname\SMALL bsy\@nd  
\scriptfont\bsyfam=\csname\smSMALL bsy\@nd 
\def\bsy{\fam\bsyfam\csname\SMALL bsy\@nd}%
\textfont\bssfam=\csname\SMALL bss\@nd  
\scriptfont\bssfam=\csname\smSMALL bss\@nd 
\scriptscriptfont\bssfam=\csname\vsSMALL bss\@nd 
\def\bss{\fam\bssfam\csname\SMALL bss\@nd}%
\ifams 
\textfont\msafam=\csname\SMALL msa\@nd  
\scriptfont\msafam=\csname\smSMALL msa\@nd 
\scriptscriptfont\msafam=\csname\vsSMALL msa\@nd 
\textfont\msbfam=\csname\SMALL msb\@nd  
\scriptfont\msbfam=\csname\smSMALL msb\@nd 
\scriptscriptfont\msbfam=\csname\vsSMALL msb\@nd 
\else\fi 
\normalbaselineskip=\smallb@seline pt%
\setbox\strutbox=\hbox{\vrule height.7\normalbaselineskip  
depth.3\baselineskip width0pt}%
\everymath{}%
\def\sc{\csname\smSMALL rm\@nd}\normalbaselines\rm}%
\everydisplay{\indenteddisplay 
   \gdef\labeltype{\eqlabel}}%
%
%
\def\hexnumber@#1{\ifcase#1 0\or 1\or 2\or 3\or 4\or 5\or 6\or 7\or 8\or 
 9\or A\or B\or C\or D\or E\or F\fi} 
\edef\bffam@{\hexnumber@\bffam} 
\edef\bdifam@{\hexnumber@\bdifam} 
\edef\bsyfam@{\hexnumber@\bsyfam} 
\def\undefine#1{\let#1\undefined} 
\def\newsymbol#1#2#3#4#5{\let\next@\relax 
 \ifnum#2=\thr@@\let\next@\bdifam@\else 
 \ifams 
 \ifnum#2=\@ne\let\next@\msafam@\else 
 \ifnum#2=\tw@\let\next@\msbfam@\fi\fi 
 \fi\fi 
 \mathchardef#1="#3\next@#4#5} 
\def\mathhexbox@#1#2#3{\relax 
 \ifmmode\mathpalette{}{\m@th\mathchar"#1#2#3}%
 \else\leavevmode\hbox{$\m@th\mathchar"#1#2#3$}\fi} 

\def\bi#1{{\fam\bdifam\relax#1}} 
%
%
\ifams\input amsmacro\fi 
%
%
\newsymbol\bitGamma 3000 
\newsymbol\bitDelta 3001 
\newsymbol\bitTheta 3002 
\newsymbol\bitLambda 3003 
\newsymbol\bitXi 3004 
\newsymbol\bitPi 3005 
\newsymbol\bitSigma 3006 
\newsymbol\bitUpsilon 3007 
\newsymbol\bitPhi 3008 
\newsymbol\bitPsi 3009 
\newsymbol\bitOmega 300A 
\newsymbol\balpha 300B 
\newsymbol\bbeta 300C 
\newsymbol\bgamma 300D 
\newsymbol\bdelta 300E 
\newsymbol\bepsilon 300F 
\newsymbol\bzeta 3010 
\newsymbol\bfeta 3011 
\newsymbol\btheta 3012 
\newsymbol\biota 3013 
\newsymbol\bkappa 3014 
\newsymbol\blambda 3015 
\newsymbol\bmu 3016 
\newsymbol\bnu 3017 
\newsymbol\bxi 3018 
\newsymbol\bpi 3019 
\newsymbol\brho 301A 
\newsymbol\bsigma 301B 
\newsymbol\btau 301C 
\newsymbol\bupsilon 301D 
\newsymbol\bphi 301E 
\newsymbol\bchi 301F 
\newsymbol\bpsi 3020 
\newsymbol\bomega 3021 
\newsymbol\bvarepsilon 3022 
\newsymbol\bvartheta 3023 
\newsymbol\bvaromega 3024 
\newsymbol\bvarrho 3025 
\newsymbol\bvarzeta 3026 
\newsymbol\bvarphi 3027 
\newsymbol\bpartial 3040 
\newsymbol\bell 3060 
\newsymbol\bimath 307B 
\newsymbol\bjmath 307C 
\mathchardef\binfty "0\bsyfam@31 
\mathchardef\bnabla "0\bsyfam@72 
\mathchardef\bdot "2\bsyfam@01 
\mathchardef\bGamma "0\bffam@00 
\mathchardef\bDelta "0\bffam@01 
\mathchardef\bTheta "0\bffam@02 
\mathchardef\bLambda "0\bffam@03 
\mathchardef\bXi "0\bffam@04 
\mathchardef\bPi "0\bffam@05 
\mathchardef\bSigma "0\bffam@06 
\mathchardef\bUpsilon "0\bffam@07 
\mathchardef\bPhi "0\bffam@08 
\mathchardef\bPsi "0\bffam@09 
\mathchardef\bOmega "0\bffam@0A 
\mathchardef\itGamma "0100 
\mathchardef\itDelta "0101 
\mathchardef\itTheta "0102 
\mathchardef\itLambda "0103 
\mathchardef\itXi "0104 
\mathchardef\itPi "0105 
\mathchardef\itSigma "0106 
\mathchardef\itUpsilon "0107 
\mathchardef\itPhi "0108 
\mathchardef\itPsi "0109 
\mathchardef\itOmega "010A 
\mathchardef\Gamma "0000 
\mathchardef\Delta "0001 
\mathchardef\Theta "0002 
\mathchardef\Lambda "0003 
\mathchardef\Xi "0004 
\mathchardef\Pi "0005 
\mathchardef\Sigma "0006 
\mathchardef\Upsilon "0007 
\mathchardef\Phi "0008 
\mathchardef\Psi "0009 
\mathchardef\Omega "000A 
%
%
\newcount\firstpage  \firstpage=1  
\newcount\jnl                      
\newcount\secno                    
\newcount\subno                    
\newcount\subsubno                 
\newcount\appno                    
\newcount\tabno                    
\newcount\figno                    
\newcount\countno                  
\newcount\refno                    
\newcount\eqlett     \eqlett=97    
\newif\ifletter 
\newif\ifwide 
\newif\ifnotfull 
\newif\ifaligned 
\newif\ifnumbysec   
\newif\ifappendix 
\newif\ifnumapp 
\newif\ifssf 
\newif\ifppt 
\newdimen\t@bwidth 
\newdimen\c@pwidth 
\newdimen\digitwidth                    
\newdimen\argwidth                      
\newdimen\secindent    \secindent=5pc   
\newdimen\textind    \textind=16pt      
\newdimen\tempval                       
\newskip\beforesecskip 
\def\beforesecspace{\vskip\beforesecskip\relax} 
\newskip\beforesubskip 
\def\beforesubspace{\vskip\beforesubskip\relax} 
\newskip\beforesubsubskip 
\def\beforesubsubspace{\vskip\beforesubsubskip\relax} 
\newskip\secskip 
\def\secspace{\vskip\secskip\relax} 
\newskip\subskip 
\def\subspace{\vskip\subskip\relax} 
\newskip\insertskip 
\def\insertspace{\vskip\insertskip\relax} 
\def\sp@ce{\ifx\next*\let\next=\@ssf 
               \else\let\next=\@nossf\fi\next} 
\def\@ssf#1{\nobreak\secspace\global\ssftrue\nobreak} 
\def\@nossf{\nobreak\secspace\nobreak\noindent\ignorespaces} 
\def\subsp@ce{\ifx\next*\let\next=\@sssf 
               \else\let\next=\@nosssf\fi\next} 
\def\@sssf#1{\nobreak\subspace\global\ssftrue\nobreak} 
\def\@nosssf{\nobreak\subspace\nobreak\noindent\ignorespaces} 
\beforesecskip=24pt plus12pt minus8pt 
\beforesubskip=12pt plus6pt minus4pt 
\beforesubsubskip=12pt plus6pt minus4pt 
\secskip=12pt plus 2pt minus 2pt 
\subskip=6pt plus3pt minus2pt 
\insertskip=18pt plus6pt minus6pt%
\fontdimen16\tensy=2.7pt 
\fontdimen17\tensy=2.7pt 
%
%
\def\eqlabel{(\ifappendix\applett 
               \ifnumbysec\ifnum\secno>0 \the\secno\fi.\fi 
               \else\ifnumbysec\the\secno.\fi\fi\the\countno)} 
\def\seclabel{\ifappendix\ifnumapp\else\applett\fi 
    \ifnum\secno>0 \the\secno 
    \ifnumbysec\ifnum\subno>0.\the\subno\fi\fi\fi 
    \else\the\secno\fi\ifnum\subno>0.\the\subno 
         \ifnum\subsubno>0.\the\subsubno\fi\fi} 
\def\tablabel{\ifappendix\applett\fi\the\tabno} 
\def\figlabel{\ifappendix\applett\fi\the\figno} 
\def\gac{\global\advance\countno by 1} 
%
%
 
\def\vfootnote#1{\insert\footins\bgroup 
\interlinepenalty=\interfootnotelinepenalty 
\splittopskip=\ht\strutbox 
\splitmaxdepth=\dp\strutbox \floatingpenalty=20000 
\leftskip=0pt \rightskip=0pt \spaceskip=0pt \xspaceskip=0pt%
\noindent\smallfonts\rm #1\ \ignorespaces\footstrut\futurelet\next\fo@t} 
%
%
\def\endinsert{\egroup 
    \if@mid \dimen@=\ht0 \advance\dimen@ by\dp0 
       \advance\dimen@ by12\p@ \advance\dimen@ by\pagetotal 
       \ifdim\dimen@>\pagegoal \@midfalse\p@gefalse\fi\fi 
    \if@mid \insertspace \box0 \par \ifdim\lastskip<\insertskip 
    \removelastskip \penalty-200 \insertspace \fi 
    \else\insert\topins{\penalty100 
       \splittopskip=0pt \splitmaxdepth=\maxdimen  
       \floatingpenalty=0 
       \ifp@ge \dimen@=\dp0 
       \vbox to\vsize{\unvbox0 \kern-\dimen@}%
       \else\box0\nobreak\insertspace\fi}\fi\endgroup}    
%
%
%
\def\ind{\hbox to \secindent{\hfill}} 
%
%

%
%
 
%
%
\def\indeqn#1{\alignedfalse\displ@y\halign{\hbox to \displaywidth 
    {$\ind\@lign\displaystyle##\hfil$}\crcr #1\crcr}} 
%
%
\def\indalign#1{\alignedtrue\displ@y \tabskip=0pt  
  \halign to\displaywidth{\ind$\@lign\displaystyle{##}$\tabskip=0pt 
    &$\@lign\displaystyle{{}##}$\hfill\tabskip=\centering 
    &\llap{$\@lign\hbox{\rm##}$}\tabskip=0pt\crcr 
    #1\crcr}} 
\def\fl{{\hskip-\secindent}} 
\def\indenteddisplay#1$${\indispl@y{#1 }} 
\def\indispl@y#1{\disptest#1\eqalignno\eqalignno\disptest} 
\def\disptest#1\eqalignno#2\eqalignno#3\disptest{%
    \ifx#3\eqalignno 
    \indalign#2%
    \else\indeqn{#1}\fi$$} 
%
%
 
%
%
 
%
%
 
%
%
 
%
%
\def\ms{\noalign{\vskip3pt plus3pt minus2pt}} 
 
\def\ns{\noalign{\vskip-3pt}}

%
 
%
%
\def\bhbar{\rlap{\kern1pt\raise.4ex\hbox{\bf\char'40}}\bi{h}} 

\def\d{{\rm d}} 
\def\e{{\rm e}} 
 
\def\frac#1#2{{#1\over#2}} 
\ifams 
\def\lap{\lesssim} 
\def\gap{\gtrsim}

\let\leq=\leqslant

\let\geq=\geqslant 
\else

\def\gap{\;\lower3pt\hbox{$\buildrel > \over \sim$}\;}%
\def\lap{\;\lower3pt\hbox{$\buildrel < \over \sim$}\;}\fi 
 
\chardef\ii="10 
\def\tqs{\hbox to 25pt{\hfil}}

\def\Bbbone{1\kern-.22em {\rm l}} 
%
%
\def\rp{\raise8pt\hbox{$\scriptstyle\prime$}} 
%
%
%
%

%
%
\def\[#1\]{\setbox0=\hbox{$\dsty#1$}\argwidth=\wd0 
    \setbox0=\hbox{$\left[\box0\right]$}\advance\argwidth by -\wd0 
    \left[\kern.3\argwidth\box0\kern.3\argwidth\right]} 
%
%
\def\lsb#1\rsb{\setbox0=\hbox{$#1$}\argwidth=\wd0 
    \setbox0=\hbox{$\left[\box0\right]$}\advance\argwidth by -\wd0 
    \left[\kern.3\argwidth\box0\kern.3\argwidth\right]} 
%
 
%
%
 
%
\def\pt(#1){({\it #1\/})} 
\let\dsty=\displaystyle

%
%
\def\reactions#1{\vskip 12pt plus2pt minus2pt%
\vbox{\hbox{\kern\secindent\vrule\kern12pt%
\vbox{\kern0.5pt\vbox{\hsize=24pc\parindent=0pt\smallfonts\rm NUCLEAR  
REACTIONS\strut\quad #1\strut}\kern0.5pt}\kern12pt\vrule}}} 
%
%
\def\slashchar#1{\setbox0=\hbox{$#1$}\dimen0=\wd0%
\setbox1=\hbox{/}\dimen1=\wd1%
\ifdim\dimen0>\dimen1%
\rlap{\hbox to \dimen0{\hfil/\hfil}}#1\else                                         
\rlap{\hbox to \dimen1{\hfil$#1$\hfil}}/\fi} 
%
%
\def\textindent#1{\noindent\hbox to \parindent{#1\hss}\ignorespaces} 
%
%
\def\opencirc{\raise1pt\hbox{$\scriptstyle{\bigcirc}$}} 
 
\ifams 
\def\opensqr{\hbox{$\square$}} 
 
\def\opentridown{\hbox{$\triangledown$}}

\else 
\def\opensqr{\vbox{\hrule height.4pt\hbox{\vrule width.4pt height3.5pt 
    \kern3.5pt\vrule width.4pt}\hrule height.4pt}} 
 
\def\opentridown{\raise1pt\hbox{$\scriptstyle\bigtriangledown$}}

\fi

%
%
\def\m@th{\mathsurround=0pt} 
%
%
\def\cases#1{%
\left\{\,\vcenter{\normalbaselines\openup1\jot\m@th%
     \ialign{$\displaystyle##\hfil$&\rm\tqs##\hfil\crcr#1\crcr}}\right.}%
%
%
\def\oldcases#1{\left\{\,\vcenter{\normalbaselines\m@th 
    \ialign{$##\hfil$&\rm\quad##\hfil\crcr#1\crcr}}\right.} 
%
%
\def\numcases#1{\left\{\,\vcenter{\baselineskip=15pt\m@th%
     \ialign{$\displaystyle##\hfil$&\rm\tqs##\hfil 
     \crcr#1\crcr}}\right.\hfill 
     \vcenter{\baselineskip=15pt\m@th%
     \ialign{\rlap{$\phantom{\displaystyle##\hfil}$}\tabskip=0pt&\en 
     \rlap{\phantom{##\hfil}}\crcr#1\crcr}}} 
\def\ptnumcases#1{\left\{\,\vcenter{\baselineskip=15pt\m@th%
     \ialign{$\displaystyle##\hfil$&\rm\tqs##\hfil 
     \crcr#1\crcr}}\right.\hfill 
     \vcenter{\baselineskip=15pt\m@th%
     \ialign{\rlap{$\phantom{\displaystyle##\hfil}$}\tabskip=0pt&\enpt 
     \rlap{\phantom{##\hfil}}\crcr#1\crcr}}\global\eqlett=97 
     \global\advance\countno by 1} 
%
%
\def\eq(#1){\ifaligned\@mp(#1)\else\hfill\llap{{\rm (#1)}}\fi} 
\def\ceq(#1){\ns\ns\ifaligned\@mp\fi\eq(#1)\cr\ns\ns} 
\def\eqpt(#1#2){\ifaligned\@mp(#1{\it #2\/}) 
                    \else\hfill\llap{{\rm (#1{\it #2\/})}}\fi} 
\let\eqno=\eq 
%
%
\countno=1 
 
\def\aleq{&\rm(\ifappendix\applett 
               \ifnumbysec\ifnum\secno>0 \the\secno\fi.\fi 
               \else\ifnumbysec\the\secno.\fi\fi\the\countno} 
\def\noaleq{\hfill\llap\bgroup\rm(\ifappendix\applett 
               \ifnumbysec\ifnum\secno>0 \the\secno\fi.\fi 
               \else\ifnumbysec\the\secno.\fi\fi\the\countno} 
\def\@mp{&} 
\def\en{\ifaligned\aleq)\else\noaleq)\egroup\fi\gac} 
\def\cen{\ns\ns\ifaligned\@mp\fi\en\cr\ns\ns} 
\def\enpt{\ifaligned\aleq{\it\char\the\eqlett})\else 
    \noaleq{\it\char\the\eqlett})\egroup\fi 
    \global\advance\eqlett by 1} 
\def\endpt{\ifaligned\aleq{\it\char\the\eqlett})\else 
    \noaleq{\it\char\the\eqlett})\egroup\fi 
    \global\eqlett=97\gac} 
%
%

\def\JPA{{\it J. Phys. A: Math. Gen.}} 

 

%
%

\def\JP{{\it J. Physique\/}}

\def\PR{{\it Phys. Rev.}} 
\def\PRL{{\it Phys. Rev. Lett.}}

\def\ZP{{\it Z. Phys.}} 
\headline={\ifodd\pageno{\ifnum\pageno=\firstpage\hfill 
   \else\rrhead\fi}\else\lrhead\fi} 
\def\rrhead{\textfonts\hskip\secindent\it 
    \shorttitle\hfill\rm\folio} 
\def\lrhead{\textfonts\hbox to\secindent{\rm\folio\hss}%
    \it\aunames\hss} 
\footline={\ifnum\pageno=\firstpage \hfill\textfonts\rm\folio\fi} 
\def\@rticle#1#2{\vglue.5pc 
    {\parindent=\secindent \bf #1\par} 
     \vskip2.5pc 
    {\exhyphenpenalty=10000\hyphenpenalty=10000 
     \baselineskip=18pt\raggedright\noindent 
     \headfonts\bf#2\par}\futurelet\next\sh@rttitle}%
\def\title#1{\gdef\shorttitle{#1} 
    \vglue4pc{\exhyphenpenalty=10000\hyphenpenalty=10000  
    \baselineskip=18pt  
    \raggedright\parindent=0pt 
    \headfonts\bf#1\par}\futurelet\next\sh@rttitle}  

\def\article#1#2{\gdef\shorttitle{#2}\@rticle{#1}{#2}}  
\def\review#1{\gdef\shorttitle{#1}%
    \@rticle{REVIEW \ifpbm\else ARTICLE\fi}{#1}} 
\def\topical#1{\gdef\shorttitle{#1}%
    \@rticle{TOPICAL REVIEW}{#1}} 
\def\comment#1{\gdef\shorttitle{#1}%
    \@rticle{COMMENT}{#1}} 
\def\note#1{\gdef\shorttitle{#1}%
    \@rticle{NOTE}{#1}} 
\def\prelim#1{\gdef\shorttitle{#1}%
    \@rticle{PRELIMINARY COMMUNICATION}{#1}} 
\def\letter#1{\gdef\shorttitle{Letter to the Editor}%
     \gdef\aunames{Letter to the Editor} 
     \global\lettertrue\ifnum\jnl=7\global\letterfalse\fi 
     \@rticle{LETTER TO THE EDITOR}{#1}} 
\def\sh@rttitle{\ifx\next[\let\next=\sh@rt 
                \else\let\next=\f@ll\fi\next} 
\def\sh@rt[#1]{\gdef\shorttitle{#1}} 
\def\f@ll{} 
\def\author#1{\ifletter\else\gdef\aunames{#1}\fi\vskip1.5pc 
    {\parindent=\secindent   
     \hang\textfonts   
     \ifppt\bf\else\rm\fi#1\par}   
     \ifppt\bigskip\else\smallskip\fi 
     \futurelet\next\@unames} 
\def\@unames{\ifx\next[\let\next=\short@uthor 
                 \else\let\next=\@uthor\fi\next} 
\def\short@uthor[#1]{\gdef\aunames{#1}} 
\def\@uthor{} 
\def\address#1{{\parindent=\secindent 
    \exhyphenpenalty=10000\hyphenpenalty=10000 
\ifppt\textfonts\else\smallfonts\fi\hang\raggedright\rm#1\par}%
    \ifppt\bigskip\fi} 
\def\jl#1{\global\jnl=#1} 
\jl{0}%
\def\journal{\ifnum\jnl=1 J. Phys.\ A: Math.\ Gen.\  
        \else\ifnum\jnl=2 J. Phys.\ B: At.\ Mol.\ Opt.\ Phys.\  
        \else\ifnum\jnl=3 J. Phys.:\ Condens. Matter\  
        \else\ifnum\jnl=4 J. Phys.\ G: Nucl.\ Part.\ Phys.\  
        \else\ifnum\jnl=5 Inverse Problems\  
        \else\ifnum\jnl=6 Class. Quantum Grav.\  
        \else\ifnum\jnl=7 Network\  
        \else\ifnum\jnl=8 Nonlinearity\ 
        \else\ifnum\jnl=9 Quantum Opt.\ 
        \else\ifnum\jnl=10 Waves in Random Media\ 
        \else\ifnum\jnl=11 Pure Appl. Opt.\  
        \else\ifnum\jnl=12 Phys. Med. Biol.\ 
        \else\ifnum\jnl=13 Modelling Simulation Mater.\ Sci.\ Eng.\  
        \else\ifnum\jnl=14 Plasma Phys. Control. Fusion\  
        \else\ifnum\jnl=15 Physiol. Meas.\  
        \else\ifnum\jnl=16 Sov.\ Lightwave Commun.\ 
        \else\ifnum\jnl=17 J. Phys.\ D: Appl.\ Phys.\ 
        \else\ifnum\jnl=18 Supercond.\ Sci.\ Technol.\ 
        \else\ifnum\jnl=19 Semicond.\ Sci.\ Technol.\ 
        \else\ifnum\jnl=20 Nanotechnology\ 
        \else\ifnum\jnl=21 Meas.\ Sci.\ Technol.\  
        \else\ifnum\jnl=22 Plasma Sources Sci.\ Technol.\  
        \else\ifnum\jnl=23 Smart Mater.\ Struct.\  
        \else\ifnum\jnl=24 J.\ Micromech.\ Microeng.\ 
   \else Institute of Physics Publishing\  
   \fi\fi\fi\fi\fi\fi\fi\fi\fi\fi\fi\fi\fi\fi\fi 
   \fi\fi\fi\fi\fi\fi\fi\fi\fi} 
\let\abs=\beginabstract 

\let\endabs=\endabstract 
\def\submitted{\ifppt\noindent\textfonts\rm Submitted to \journal\par 
     \bigskip\fi} 
\def\today{\number\day\ \ifcase\month\or 
     January\or February\or March\or April\or May\or June\or 
     July\or August\or September\or October\or November\or 
     December\fi\space \number\year} 
\def\date{\ifppt\noindent\textfonts\rm  
     Date: \today\par\goodbreak\bigskip\fi} 
%
%
\def\pacs#1{\ifppt\noindent\textfonts\rm  
     PACS number(s): #1\par\bigskip\fi} 
%
 
%
%
\def\section#1{\ifppt\ifnum\secno=0\eject\fi\fi 
    \subno=0\subsubno=0\global\advance\secno by 1 
    \gdef\labeltype{\seclabel}\ifnumbysec\countno=1\fi 
    \goodbreak\beforesecspace\nobreak 
    \noindent{\bf \the\secno. #1}\par\futurelet\next\sp@ce} 
\def\subsection#1{\subsubno=0\global\advance\subno by 1 
     \gdef\labeltype{\seclabel}%
     \ifssf\else\goodbreak\beforesubspace\fi 
     \global\ssffalse\nobreak 
     \noindent{\it \the\secno.\the\subno. #1\par}%
     \futurelet\next\subsp@ce} 
\def\subsubsection#1{\global\advance\subsubno by 1 
     \gdef\labeltype{\seclabel}%
     \ifssf\else\goodbreak\beforesubsubspace\fi 
     \global\ssffalse\nobreak 
     \noindent{\it \the\secno.\the\subno.\the\subsubno. #1}\null.  
     \ignorespaces} 
%
 
%
%
\def\numappendix#1{\ifappendix\ifnumbysec\countno=1\fi\else 
    \countno=1\figno=0\tabno=0\fi 
    \subno=0\global\advance\appno by 1 
    \secno=\appno\gdef\applett{A}\gdef\labeltype{\seclabel}%
    \global\appendixtrue\global\numapptrue 
    \goodbreak\beforesecspace\nobreak 
    \noindent{\bf Appendix \the\appno. #1\par}%
    \futurelet\next\sp@ce} 
\def\numsubappendix#1{\global\advance\subno by 1\subsubno=0 
    \gdef\labeltype{\seclabel}%
    \ifssf\else\goodbreak\beforesubspace\fi 
    \global\ssffalse\nobreak 
    \noindent{\it A\the\appno.\the\subno. #1\par}%
    \futurelet\next\subsp@ce} 
\def\@ppendix#1#2#3{\countno=1\subno=0\subsubno=0\secno=0\figno=0\tabno=0 
    \gdef\applett{#1}\gdef\labeltype{\seclabel}\global\appendixtrue 
    \goodbreak\beforesecspace\nobreak 
    \noindent{\bf Appendix#2#3\par}\futurelet\next\sp@ce} 
\def\Appendix#1{\@ppendix{A}{. }{#1}} 
\def\appendix#1#2{\@ppendix{#1}{ #1. }{#2}} 
\def\App#1{\@ppendix{A}{ }{#1}} 
\def\app{\@ppendix{A}{}{}} 
\def\subappendix#1#2{\global\advance\subno by 1\subsubno=0 
    \gdef\labeltype{\seclabel}%
    \ifssf\else\goodbreak\beforesubspace\fi 
    \global\ssffalse\nobreak 
    \noindent{\it #1\the\subno. #2\par}%
    \nobreak\subspace\noindent\ignorespaces} 
%
%
\def\@ck#1{\ifletter\bigskip\noindent\ignorespaces\else 
    \goodbreak\beforesecspace\nobreak 
    \noindent{\bf Acknowledgment#1\par}%
    \nobreak\secspace\noindent\ignorespaces\fi} 
\def\ack{\@ck{s}} 
\def\ackn{\@ck{}} 
\def\n@ip#1{\goodbreak\beforesecspace\nobreak 
    \noindent\smallfonts{\it #1}. \rm\ignorespaces} 
\def\naip{\n@ip{Note added in proof}} 
\def\na{\n@ip{Note added}} 
 
%
%
 
%
 
%
%
 
%
 
%
 
\def\tablecont{\topinsert\global\advance\tabno by -1 
    \tablecaption{(continued)}} 
\def\tablecaption#1{\gdef\labeltype{\tablabel}\global\widefalse 
    \leftskip=\secindent\parindent=0pt 
    \global\advance\tabno by 1 
    \smallfonts{\bf Table \ifappendix\applett\fi\the\tabno.} \rm #1\par 
    \smallskip\futurelet\next\t@b} 
\def\t@b{\ifx\next*\let\next=\widet@b 
             \else\ifx\next[\let\next=\fullwidet@b 
                      \else\let\next=\narrowt@b\fi\fi 
             \next} 
\def\widet@b#1{\global\widetrue\global\notfulltrue 
    \t@bwidth=\hsize\advance\t@bwidth by -\secindent}  
\def\fullwidet@b[#1]{\global\widetrue\global\notfullfalse 
    \leftskip=0pt\t@bwidth=\hsize}                   
\def\narrowt@b{\global\notfulltrue} 
\def\align{\catcode`?=13\ifnotfull\moveright\secindent\fi 
    \vbox\bgroup\halign\ifwide to \t@bwidth\fi 
    \bgroup\strut\tabskip=1.2pc plus1pc minus.5pc} 
\def\endalign{\egroup\egroup\catcode`?=12} 
 
%
%

%
%

%
 
%
%

%
 
\catcode`?=13 
\def\lineup{\setbox0=\hbox{\smallfonts\rm 0}%
    \digitwidth=\wd0%
    \def?{\kern\digitwidth}%
    \def\\{\hbox{$\phantom{-}$}}%
    \def\-{\llap{$-$}}} 
\catcode`?=12 
%
%
\def\sidetable#1#2{\hbox{\ifppt\hsize=18pc\t@bwidth=18pc 
                          \else\hsize=15pc\t@bwidth=15pc\fi 
    \parindent=0pt\vtop{\null #1\par}%
    \ifppt\hskip1.2pc\else\hskip1pc\fi 
    \vtop{\null #2\par}}}  
\def\lstable#1#2{\everypar{}\tempval=\hsize\hsize=\vsize 
    \vsize=\tempval\hoffset=-3pc 
    \global\tabno=#1\gdef\labeltype{\tablabel}%
    \noindent\smallfonts{\bf Table \ifappendix\applett\fi 
    \the\tabno.} \rm #2\par 
    \smallskip\futurelet\next\t@b} 
\def\inctabno{\global\advance\tabno by 1} 
%
%
 
%
 
%
\def\figure#1{\figc@ption{#1}\bigskip} 
\def\figc@ption#1{\global\advance\figno by 1\gdef\labeltype{\figlabel}%
   {\parindent=\secindent\smallfonts\hang 
    {\bf Figure \ifappendix\applett\fi\the\figno.} \rm #1\par}} 
%
%
\def\refHEAD{\goodbreak\beforesecspace 
     \noindent\textfonts{\bf References}\par 
     \let\ref=\rf 
     \nobreak\smallfonts\rm} 
\def\references{\refHEAD\parindent=0pt 
     \everypar{\hangindent=18pt\hangafter=1 
     \frenchspacing\rm}%
     \secspace} 
\def\rf#1{\par\noindent\hbox to 21pt{\hss #1\quad}\ignorespaces} 
%
 
%
 
%
%
\def\numrefjl#1#2#3#4#5{\par\rf{#1}#2 {\it #3 \bf #4} #5\par} 
%
%
\def\numrefbk#1#2#3#4{\par\rf{#1}#2 {\it #3} #4\par} 
%
%

%
%

%
\catcode`\@=12 
%
%
 
%
%
\def\jnlstyle{\pptfalse\headsize{14}{18}%
\textsize{10}{12}%
\smallsize{8}{10} 
\textind=16pt} 
%
%
 
%
%
 
%
\parindent=\textind 
%
\input epsf
\def\received#1{\insertspace 
     \parindent=\secindent\ifppt\textfonts\else\smallfonts\fi 
     \hang{Received #1}\rm } 
\def\figure#1{\global\advance\figno by 1\gdef\labeltype{\figlabel}%
   {\parindent=\secindent\smallfonts\hang 
    {\bf Figure \ifappendix\applett\fi\the\figno.} \rm #1\par}} 
\headline={\ifodd\pageno{\ifnum\pageno=\firstpage\titlehead
   \else\rrhead\fi}\else\lrhead\fi} 
\def\lpsn#1#2{LPSN-#1-LT#2}

\footline={\ifnum\pageno=\firstpage{\smallfonts cond-mat/9411077}
\hfil\textfonts\rm\folio\fi}   

\def\titlehead{\smallfonts J. Phys. A: Math. Gen.  {\bf 28} (1995) 351--363 
\hfil\lpsn{94}{5}} 

\firstpage=351
\pageno=351

\jnlstyle
\jl{1}
\overfullrule=0pt

\title{Surface shape and local critical behaviour\hfill\break in two-dimensional
directed percolation}[Surface shape and local critical behaviour]
 
\author{C Kaiser and L Turban}[C Kaiser and L Turban]
 
\address{Laboratoire de Physique du Solide\footnote\dag{Unit\'e de
Recherche Associ\'ee au CNRS No 155},  Universit\'e Henri Poincar\'e (Nancy~I),
BP 239, F--54506 Vand\oe uvre l\`es Nancy Cedex, France}

\received{23 September 1994}

\abs
Two--dimensional directed site
percolation is studied in systems directed along the $\scriptstyle x$--axis and
limited by a free surface at $\scriptstyle y=\pm Cx^k$. Scaling
considerations show that the surface is a re\-le\-vant per\-tur\-ba\-tion to the
local critical behaviour when $\scriptstyle k<1/z$ where
$\scriptstyle z=\nu_\parallel/\nu$ is the dynamical exponent. The
tip--to--bulk order parameter correlation function is calculated in the
mean--field approximation. The tip percolation probability and the fractal
dimensions of critical clusters are obtained through Monte--Carlo simulations.
The tip order parameter has a nonuniversal, $\scriptstyle C$--dependent, scaling
dimension  in the marginal case, $\scriptstyle k=1/z$, and displays a
stretched exponential behaviour when the per\-tur\-ba\-tion is relevant. The
$\scriptstyle k$--dependence of the fractal dimensions in the relevant case is
in agreement with the results of a blob picture approach.
\endabs

\pacs{68.35.Rh, 64.60.Fr, 64.60.Ak, 64.60.Ht}
\submitted
\date

\section{Introduction} The shape of the free surface
limiting a system may influence its local critical behaviour at a bulk second
order phase transition, provided the deviation from the flat surface is long
range. For corners in two dimensions, wedges or cones in three dimensions, a
marginal behaviour is obtained, with local critical exponents depending on the
opening angle, in isotropic critical systems~[1--9]. This result is linked to
the invariance of the shapes under isotropic rescaling~(see~[10] for a review). 

With parabolic shapes, the surface introduces a relevant perturbation
to the flat surface fixed point leading, locally, to stretched exponential
behaviour for the order parameter and the correlation functions~[11]. The Ising
model~[11--13], the self--avoiding--walk~[14] and ordinary percolation~[15] have
been studied for this geometry in two dimensions.

In anisotropic systems the correlation length diverges as
$\xi_\parallel\!\sim\! t^{-\nu_\parallel}$ along a time--like direction and 
as $\xi_\perp\!\sim\! t^{-\nu}$ in the transverse directions with a
dynamical exponent $z\!=\!\nu_\parallel/\nu$~[16]. Covariance under a change of
the length scales then requires two different scaling factors, $b_\parallel\!=\!
b^z$ and $b_\perp\!=\! b$~[17]. In this way, the relation between the
correlation lengths $\xi_\parallel\!\sim\!(\xi_\perp)^z$ is preserved. As a
consequence, scale invariant shapes are different from the isotropic case: they
now correspond to parabolic--like surfaces~[18]. For example, the marginal shape
for the directed walk, with $z\!=\!2$, is the true parabola in two dimensions,
a paraboloid or a parabolic wedge in three dimensions~[18,19].

In the present work, we study the two--dimensional directed site percolation
problem inside a parabolic--like system. The scaling behaviour is discussed in
section~2. The problem is solved in a mean--field  approximation in
section~3. The results of Monte--Carlo simulations are presented in section~4
and we end with a discussion in section~5.

\section{Scaling considerations}

We consider a system displaying anisotropic critical behaviour and limited by a
free surface at $y\!=\!\pm Cx^k$ in the $(x,y)$--plane where $x$ is the
time--like coordinate. Under rescaling, with $x'\!=\! x/b^z$ and $y'\!=\! y/b$,
the geometrical constant $C$ transforms according~to
$$
{1\over C'}=b^{1-zk}{1\over C}.
\eqno(2.1)
$$
When $k\!>\!1/z$, $1/C$ renormalizes to zero, i. e. the system flows towards the
flat surface geometry. The perturbation introduced by the free surface is 
irrelevant with respect to the flat surface fixed point. On the contrary, when
$k\!<\!1/z$, $1/C$ grows under rescaling and the system becomes locally
narrower. The perturbation is then relevant and one expects a strong reduction of
the order at the tip associated with a new type of local critical behaviour.
The value $k\!=\!1/z$ corresponds to a scale--invariant shape leading to a
marginal local critical behaviour with $C$--dependent exponents. For an
isotropic system, $z\!=\!1$ and the marginal shape, with $k\!=\!1$, corresponds
to the corner geometry mentioned above.

Let $m_0$ be the tip order parameter, with scaling dimension
$x_m$ at the flat surface fixed point, on a finite system  with size
$L$ along the $x$--axis. Under a change of scale, it transforms as
$$
m_0\left(t,{1\over C},{1\over L}\right)=
b^{-x_m}m_0\left(b^{1/\nu}t,{b^{1-zk}\over C},{b^z\over L}\right) 
\eqno(2.2)
$$
where $t$ is the thermal scaling field and  $1/C$, which is vanishing at the
reference fixed point like $1/L$ in finite--size scaling, is treated as a new
scaling field introduced by the parabolic free surface. One may also notice
that, for a directed problem, the system does not see a flat free surface at
$x\!=\!0$ and  $x_m$ is also the bulk scaling dimension of the order parameter.

Taking $b\!=\! t^{-\nu}$, one obtains
$$
m_0\left(t,{1\over C},{1\over L}\right)=
t^{\beta}m_t\left({l_C\over t^{-\nu_\parallel}},{L\over
t^{-\nu_\parallel}}\right),\qquad l_C=C^{z/(1-zk)}
\eqno(2.3)
$$
where $\beta\!=\!\nu x_m$ and the shape of the system introduces a new
length scale given by $l_C$ as long as the shape is not scale--invariant, i. e.
$k\!\neq\!1/z$. 
In the same way, taking $b\!=\! l_C^{1/z}$ in equation~(2.2), one obtains
the following finite--size scaling behaviour at the critical point $t\!=\!0$:
$$
m_0\left(0,{1\over C},{1\over L}\right)=l_C^{-x_m/z}m_C\left({L\over l_C}\right).
\eqno(2.4)
$$

In the marginal case, $z\!=\!1/k$, the scaling dimension of the tip order
parameter becomes $C$--dependent and equation~(2.2) is changed into
$$
m_0\left(t,{1\over C},{1\over
L}\right)=b^{-x_m(C)}m_{marg}\left(b^{1/\nu}t,{b^z\over L}\right). 
\eqno(2.5)
$$

The order parameter correlation function between the origin and a
point at $(x,y)$ transforms as
$$
G\left(t,x,y,{1\over C}\right)=b^{-2x_m}G\left(b^{1/\nu}t,{x\over b^z},{y\over
b},{b^{1-zk}\over C}\right) 
\eqno(2.6)
$$
when $L$ is infinite. With $b\!=\! x^{1/z}$, equation~(2.6) leads to:
$$
G\left(t,x,y,{1\over C}\right)=x^{-2x_m/z}g\left({x\over
t^{-\nu_\parallel}},{y^z\over x},{x\over l_C}\right).
\eqno(2.7)
$$

These scaling forms will be used later to analyze the mean--field results and
the Monte--Carlo data.
 
\section{Mean--field approximation}

We consider the site percolation problem on a square lattice, directed along
a diagonal of the squares as shown in figure 1a, with a site occupation
probability $p$. The order parameter correlation function is the probability
density $P(x,y)$ for a site at $(x,y)$ to be connected to the origin. The site at
$(x\!+\!1,y)$ will be connected too if it is occupied and if at least one of its
two nearest neighbours at $x$ are themselves connected to the origin~(figures
1b--d). In a mean--field approximation~[20] where correlations between the $P$s
on different sites are neglected, this occurs with probability  
$$
\fl\eqalign{
P(x+1,y)=&p\Bigl\{P(x,y+1)\big[1-P(x,y-1)\big]+P(x,y-1)\big[1-P(x,y+1)\big]\cr
&+P(x,y+1)P(x,y-1)\Bigr\}\cr}
\eqno(3.1)
$$
where the different terms on the right correspond to the last three diagrams
in figure~1.

{\par\begingroup\medskip
\epsfxsize=9truecm
\topinsert
\centerline{\epsfbox{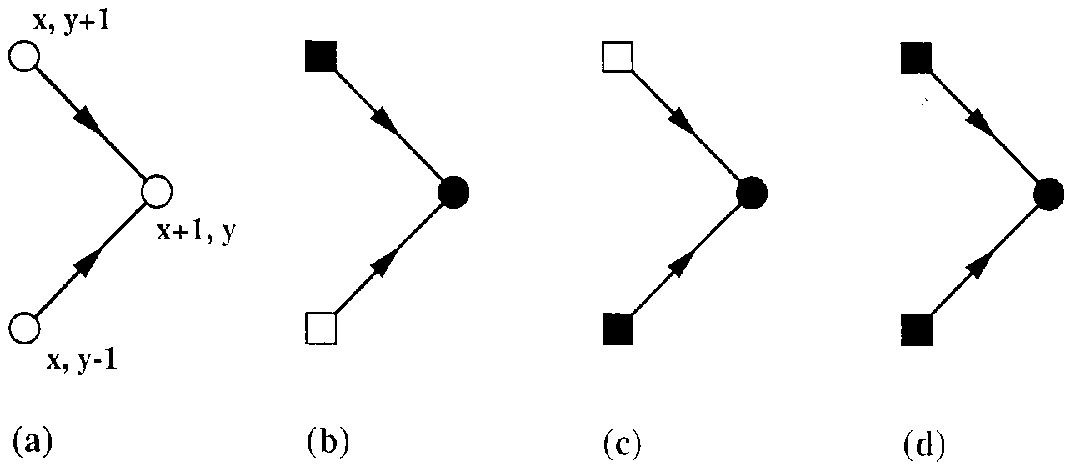}}
\smallskip
\figure{Directed site percolation on the square
lattice: (a) The system is directed along the diagonal of the square lattice.
(b--d) The site at ($x\!+\!1,y$) (full circle) is connected to the
origin if at least one of its nearest neighbours at $x$ is
connected to the origin (full squares).}
\endinsert 
\endgroup
\par}

Going to the continuum limit and neglecting nonlinear terms involving
derivatives, one is led to the following differential equation:
$$
{\partial P\over\partial x}=p{\partial^2P\over\partial y^2}+(2p-1)P-pP^2.
\eqno(3.2)
$$
The order parameter is the probability $P_0$ for the site at
the origin to belong to an infinite cluster. For the un--confined system, it is
given by the homogeneous stationary solution of equation~(3.2) as 
$$
P_0=\left\{\matrix{{2p-1\over p}&\qquad p\geq p_c={1\over 2}\cr
\ms 0&\qquad p<p_c\cr}\right.
\eqno(3.3)
$$
and vanishes at the percolation threshold $p_c$ with an exponent
$\beta^{mf}\!=\!1$. 

Below the threshold and far from the origin, $P(x,y)\ll 1$ and equation~(3.2) can
be linearized, leading to:
$$
{\partial P\over\partial x}\simeq{1\over 2}{\partial^2P\over\partial
y^2}+2(p-p_c)P. 
\eqno(3.4)
$$
Through the change of function $P(x,y)\to\exp\big[2(p_c-p)x\big]P(x,y)$ an
ordinary diffusion equation is obtained, so that
$$
P(x,y)=\e^{-2(p_c-p)x}\ {\exp\Bigl(-{y^2\over
2x}\Bigr)\over\sqrt{2\pi x}} 
\eqno(3.5)
$$
for the un--confined system. In $d\!+\!1$ dimensions the power--law decay would
be changed into $x^{-d/2}$. From a comparison with equation~(2.7) where $t$ is
now $\vert p-p_c\vert$, one deduces the following exponents for directed
percolation,  $$
\nu_\parallel^{mf}=1\qquad\nu^{mf}=1/2\qquad z^{mf}=2\qquad x_m^{mf}={d\over 2}
\eqno(3.6)
$$
where $\nu^{mf}$ follows from scaling. The scaling law $\beta=\nu x_m$ is
verified for the mean--field (Gaussian) exponents only at the upper critical
dimension $d_c+1$ which, according~to~(3.3) and~(3.6), is equal to~$5$ for
directed percolation. 

On a parabolic system, we use the new variables $x$ and $\zeta(x,y)\!=\! y/x^k$ 
for which the free surface is shifted to $\zeta\!=\!\pm C$  and equation~(3.4) is
changed into
$$
{\partial P\over\partial x}={1\over 2x^{2k}}{\partial^2P\over\partial
\zeta^2}+k{\zeta\over x}{\partial P\over\partial\zeta}+2(p-p_c)P 
\eqno(3.7)
$$
with the boundary condition $P(x,\zeta\!=\!\pm C)\!=\!0$.
Through the change of function
$$
P(x,\zeta)=\exp\big[-2(p_c-p)x-{k\over 2}\zeta^2x^{2k-1}\big]Q(x,\zeta)
\eqno(3.8)
$$
equation~(3.7) leads to
$$
{\partial Q\over\partial x}={1\over 2x^{2k}}{\partial^2 Q\over\partial
\zeta^2}+{k\over 2}\Bigl[(k-1)\zeta^2x^{2k-2}-{1\over x}\Bigr]Q  
\eqno(3.9) 
$$
for which the variables separate when $k\!=\!1,1/2$ or $0$. According to
equation~(2.1), in mean--field, these values of $k$ just correspond
to irrelevant, marginal and relevant perturbations.

For $k\!=\!1$, i. e. in the corner geometry, (3.9) gives
$$
x^2{\partial Q\over\partial x}+{x\over 2}Q={1\over 2}{\partial^2 Q\over\partial
\zeta^2}\qquad \zeta={y\over x}\qquad P=\e^{-2(p_c-p)x-{y^2/2x}} Q
\eqno(3.10)
$$
and the correlation function, which is even in $y$, takes the
form 
$$
\fl P(x,y)=\e^{-2(p_c-p)x}\ {\exp\Bigl({-y^2\over 2x}\Bigr)\over\sqrt{2\pi
x}}\ \sum_{n=0}^\infty A_n\exp\Bigl[{(2n+1)^2\pi^2\over 8C^2x}\Bigr]
\cos\Bigl[{(2n+1)\pi y\over 2C x}\Bigr] . 
\eqno(3.11)
$$
This is just the un--confined solution~(3.5), modulated by a function which
depends on the two last variables $v$ and $w$ of the scaling function~$g(u,v,w)$
in equation~(2.7), with here $z\!=\!2$ and $l_C\!=\! C^{-2}$. The critical
behaviour is the same as for un--confined percolation as expected for an
irrelevant perturbation.

For the true parabola which is the marginal geometry, one may use equation~(3.7)
with $k\!=\!1/2$ to obtain
$$
x{\partial Q\over\partial x}={1\over 2}{\partial^2 Q\over\partial
\zeta^2}+{\zeta\over 2}{\partial Q\over\partial\zeta}\qquad \zeta={y\over
x^{1/2}}\qquad P=\e^{-2(p_c-p)x} Q
\eqno(3.12)
$$
which is of the form studied in [18] for the directed walk problem. Writing
$Q(x,\zeta)\!=\!\phi(x)\psi(\zeta)$ leads to the following eigenvalue problem for
$\psi(\zeta)$ 
$$
{1\over 2}{\d^2\psi\over\d\zeta^2}+{\zeta\over
2}{\d\psi\over\d\zeta}=-\lambda^2\psi
\eqno(3.13)
$$
with $\phi(x)\!\sim\! x^{-\lambda^2}$. The solution, which is even and regular
at the origin, can be written as a series expansion, $\sum_{n=0}^\infty
a_n\zeta^{2n}$. According to~(3.13), the coefficients satisfy
$$
a_{n+1}=-{(\lambda^2+n)\over(n+1)(2n+1)}\  a_n
\eqno(3.14)
$$
which is the recursion relation for the coefficients of the degenerate
hypergeometric function
$$
\fl _1F_1\Bigl[\lambda^2,{1\over 2};-{\zeta^2\over
2}\Bigr]=1+\sum_{n=1}^\infty(-1)^n{\lambda^2(\lambda^2+1)\cdots(\lambda^2+n-1)\over
1.3.\cdots(2n-1)}\  {\zeta^{2n}\over n!} .
\eqno(3.15)
$$
The boundary condition $\psi({C})\!=\!0$ gives the eigenvalues $\lambda_n$ which
are the zeros of $_1F_1\big[\lambda_n^2,{1/2};-{C^2/2}\big]$. The solution is
obtained as the eigenvalue expansion
$$
P(x,y)=\e^{-2(p_c-p)x}\sum_{n=0}^\infty B_n\ x^{-\lambda_n^2}\
_1F_1\Bigl[\lambda_n^2,{1\over 2};-{y^2\over 2x}\Bigr] .
\eqno(3.16)
$$
The behaviour at large $x$ is governed by the first term in this
expansion which decays as $x^{-\lambda_0^2}$, i. e. with a $C$--dependent
exponent as expected for a marginal perturbation. The dimension of the
tip--to--bulk correlation function is the sum of the tip and bulk 
order parameter dimensions,  the first one being variable. Comparing with the
form of the decay in~(2.7) gives $\lambda_0^2\!=\![x_m^{mf}(C)\!+\!
x_m^{mf}]/z^{mf}$ and, using~(3.6), the tip order parameter dimension is given by
$$
x_m^{mf}(C)=2\lambda_0^2-{1\over 2} .
\eqno(3.17)
$$
Its dependence on $C$ is shown in figure~2.

{\par\begingroup\medskip
\epsfxsize=9truecm
\topinsert
\centerline{\epsfbox{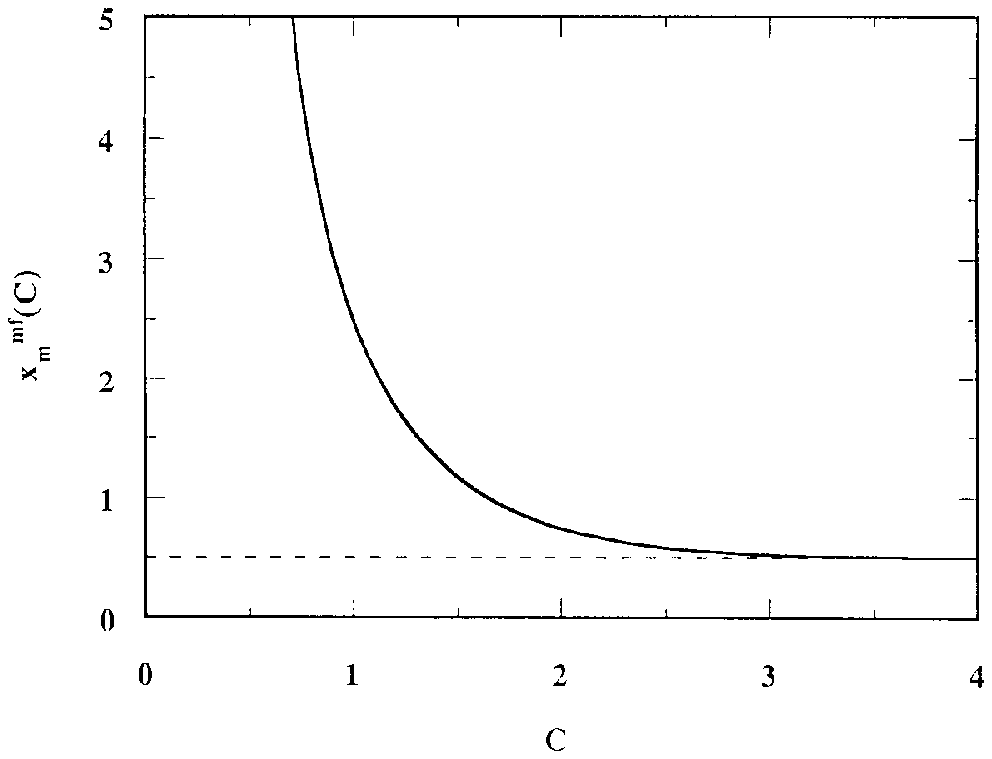}}
\smallskip
\figure{Scaling dimension of the tip order parameter
as a function of $C$ in the mean--field approximation when the
shape is marginal. The local exponent $x_m^{mf}(C)$ diverges when
$C\!\to\!0$ and goes to its unperturbed value $x_m^{mf}\!=\!1/2$ when
$C\!\to\!\infty$.} 
\endinsert 
\endgroup
\par}

Analytical results can be obtained only in limiting cases which have already been
discussed in [18]. When $C$ is infinite, $\lambda_0^2\!=\!1/2$, only the first
term in the expansion remains, which satisfies the initial and boundary
conditions, giving back the free solution in equation~(3.5) since $_1F_1({1\over
2},{1\over 2};-y^2/2x)\!=\!\exp(-y^2/2x)$. For large $C$--values, the tip
exponent takes the following form 
$$
x_m^{mf}(C)={1\over 2}+\sqrt{2\over\pi} C \exp\Bigl(-{C^2\over
2}\Bigr)[1+O(\varepsilon)] 
\eqno(3.18)
$$
where $\varepsilon$ is the correction term itself. For narrow systems, the
hypergeometric function gives a cosine to leading order in $C^2$. One obtains
the following asymptotic behaviour in $x$
$$
P(x,y)\sim x^{-\pi^2/8C^2}\cos\left({\pi y\over 2C\sqrt{x}}\right)
\eqno(3.19)
$$
and the tip exponent diverges as $\pi^2/4C^2$.

For the strip geometry, which corresponds to a relevant perturbation, one can
use~(3.9) with $k\!=\!0$ or, more simply, go back to the original equation~(3.4)
giving
$$
{\partial Q\over\partial x}={1\over 2}{\partial^2 Q\over\partial
y^2}\qquad P=\e^{-2(p_c-p)x} Q
\eqno(3.20)
$$
The solution satisfying the initial condition $P(0,y)\!=\!\delta(y)$ reads
$$ 
\fl P(x,y)={\exp\big[-2(p_c-p)x\big]\over C}\sum_{n=0}^\infty
\exp\Bigl[-{(2n+1)^2\pi^2x\over 8C^2}\Bigr] \cos\Bigl[{(2n+1)\pi
y\over 2C}\Bigr] 
\eqno(3.21)
$$
in agreement with the scaling form~(2.7) for the mean--field exponents.
The perturbation due to the free surface induces a
simple exponential decay of the correlations at $p_c$. Actually, the system is
now one--dimensional and this term corresponds to a shift of the percolation
threshold by $(\pi/4C)^2$. 

For $0\!<\! k\!<\!1/2$, the simple exponential is expected to be
replaced by a stretched one, involving some power of the
last scaling variable $w\!=\! x/l_C$ in equation~(2.7). The power goes to 1 when
$k\!=\!0$ and has to vanish when the system is marginal $(k\!=\!1/2)$ in
order to give a $C$--dependent decay exponent. 
A good candidate for the asymptotic behaviour is~[18]
$$
\fl P(x,y)={\exp\big[-2(p_c-p)x\big]\over Cx^k}\ \exp\left(-{\pi^2\over 8C^2}\
{x^{1-2k}\over 1-2k}\right)\cos\left({\pi
y\over 2Cx^k}\right)  .
\eqno(3.22)
$$
It has the proper scaling form~(2.7) with mean--field exponents and interpolates
between the strip result~(3.21) and the small--$C$ result~(3.19) for the parabola. 

\section{Monte--Carlo simulations}

As in the mean--field approach of the last section, the Monte--Carlo~(MC)
simulations were performed for site percolation on a square lattice along the
diagonal direction. Accurate estimates of the percolation threshold and 
directed percolation exponents have been deduced from series expansion in [21]
with
$$\eqalign{
&p_c=0.705489(4)\qquad\nu=1.097(2)\qquad\nu_\parallel=1.734(2)\cr
&\gamma=2.278(2)\qquad z=1.581(3)\qquad\beta=0.276(3)\cr}
\eqno(4.1)
$$ 
where the last two values follow from scaling laws.

On the lattice, the sites inside or on the curve $y\!=\!\pm Cx^k$ are
considered to belong to the system.  The tip order parameter was
calculated on parabolic--like systems with size $L$ along the $x$--axis. On
these finite--size systems, $P_0$ is the probability that a cluster grown from an
occupied site at $x\!=\! y\!=\!0$ reaches $x\!=\! L$. Such a cluster is called
"infinite". "Finite" clusters are those dying before $x\!=\! L$. 

In order to spare computer time the MC--algorithm ensures that a cluster reaches
$x\!=\! L$ as fast as possible. This is done by always trying to occupy
those sites with the largest $x$--coordinate. The coordinates of sites from
which the growth may eventually continue but which are closer to the origin
are stored into a list of active sites. The growth stops once the size of the
cluster is equal to $L$, even if it still contains active sites. Thus, the
complete infinite cluster has not been built. Subsequently, if there are no
active sites left, a complete finite cluster has been grown. An example is
shown in figure~3. 

{\par\begingroup\medskip
\epsfxsize=9truecm
\topinsert
\centerline{\epsfbox{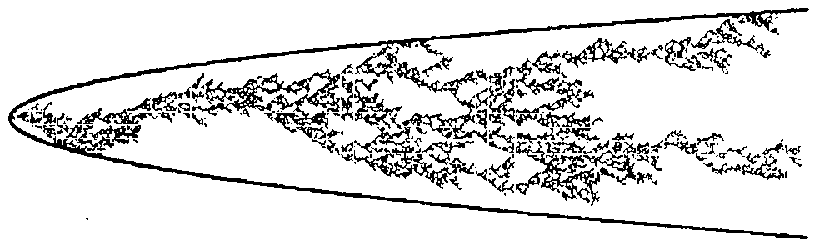}}
\smallskip
\figure{Directed percolation cluster generated using
the Monte--Carlo algorithm described in the text. Here, $L\!=\!1000$,
$k\!=\!1/2$ and $C\!=\!4$.}
\endinsert 
\endgroup
\par}

With this algorithm, which is a modified version of the one
used in [15] for ordinary percolation, the computation time for increasing
$p\!>\! p_c$ remains almost constant. Thus we could compute $P_0$ as a function
of $p$ up to $p\!=\!0.8$ $(k\!=\!1/2,1/z)$ for
$L\!=\!1000$ taking averages over $10^5$ samples. The results are shown in
figures~4 and~5 for values of~$C$ ranging from~$1$ to~$4$. 

{\par\begingroup\medskip
\epsfxsize=9truecm
\topinsert
\centerline{\epsfbox{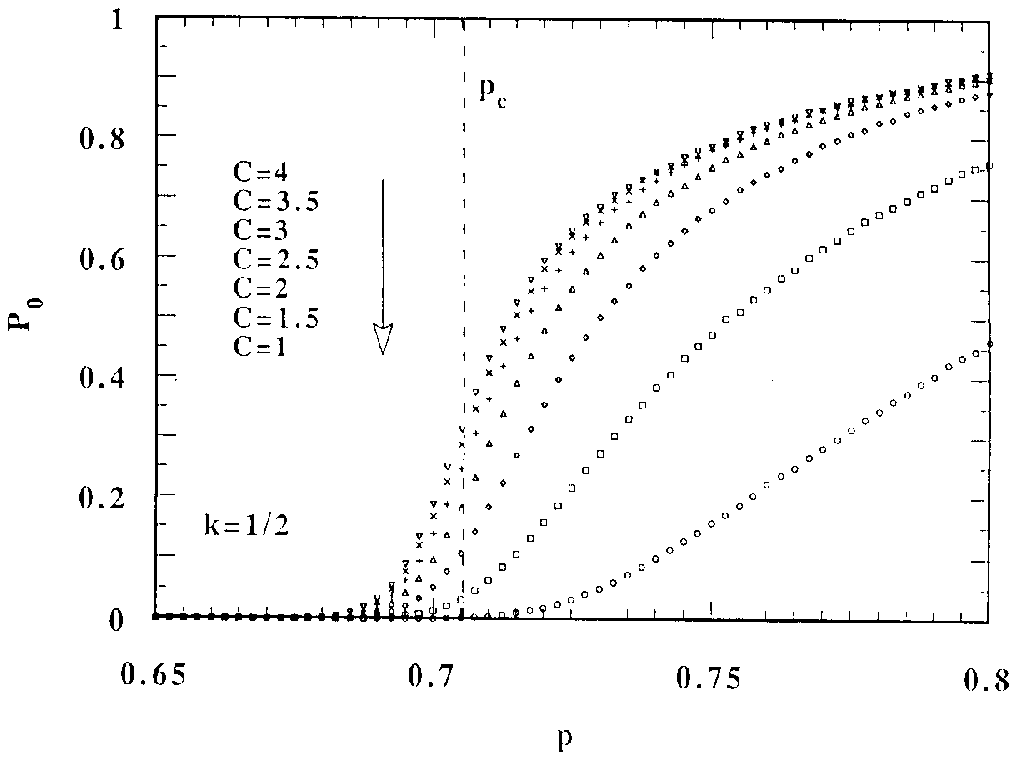}}
\smallskip
\figure{Tip percolation probability on a parabola
($k\!=\!1/2$) with length $L\!=\!1000$ as a function of
$p$. The Monte--Carlo results are averaged over $10^5$
samples for $7$ values of $C$ between $4$
and $1$ from top to bottom. The non--vanishing percolation
probability at the percolation threshold (dashed line) indicates strong
finite--size effects which are increasing with $C$.} 
\endinsert  
\endgroup
\par}

{\par\begingroup\medskip
\epsfxsize=9truecm
\topinsert
\centerline{\epsfbox{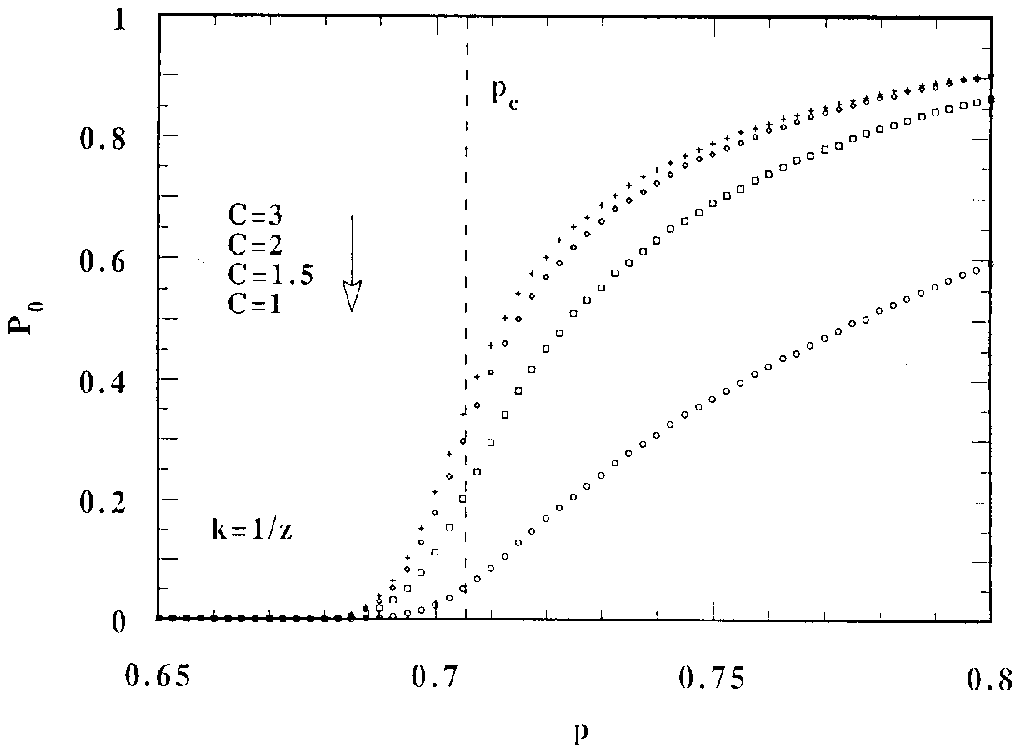}}
\smallskip
\figure{As in figure 4 for the marginal case, $k\!=\!1/z$.}
\endinsert  
\endgroup
\par}

{\par\begingroup\medskip
\epsfxsize=9truecm
\topinsert
\centerline{\epsfbox{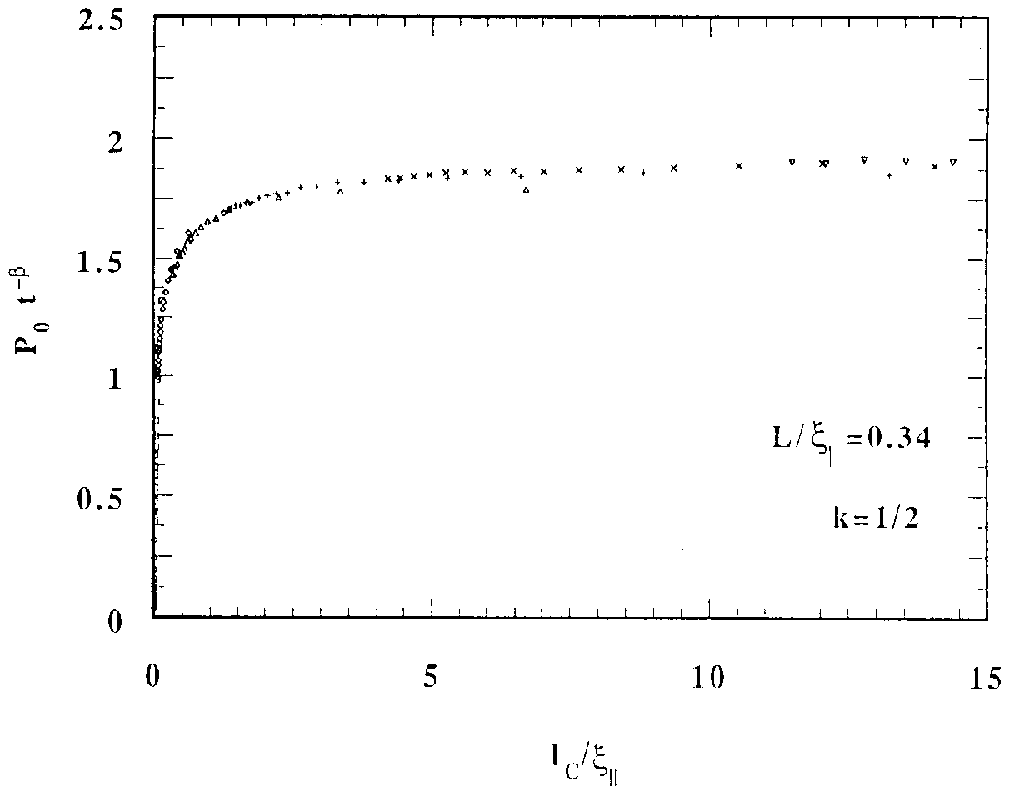}}
\smallskip
\figure{Scaling function $P_0t^{-\beta}$
versus $l_C/\xi_\parallel$ in the relevant case ($k\!=\!1/2$). A good data
collapse is obtained for the different values of $C$ (same symbols as in 
figure~4).}
\endinsert  
\endgroup \par}

Due to the anisotropy, finite--size effects remain important. This appears in the
non--vanishing values of $P_0(p_c)$ and increases with $C$, i. e. when the
system opens. According to the scaling relation~(2.3), $P_0t^{-\beta}$ is an
universal function of $l_C/\xi_\parallel$ and $L/\xi_\parallel$. This has been
verified in the relevant case $k\!=\!1/2$ for the first of these scaling
variables. A good data collapse is obtained in figure~6 for $C\!=\!1,
1.5,\dots,4$ and $L\!=\!50,100,\dots,1000$ with a fixed ratio
$L/\xi_\parallel\!=\!0.34$.

{\par\begingroup\medskip
\epsfxsize=9truecm
\topinsert
\centerline{\epsfbox{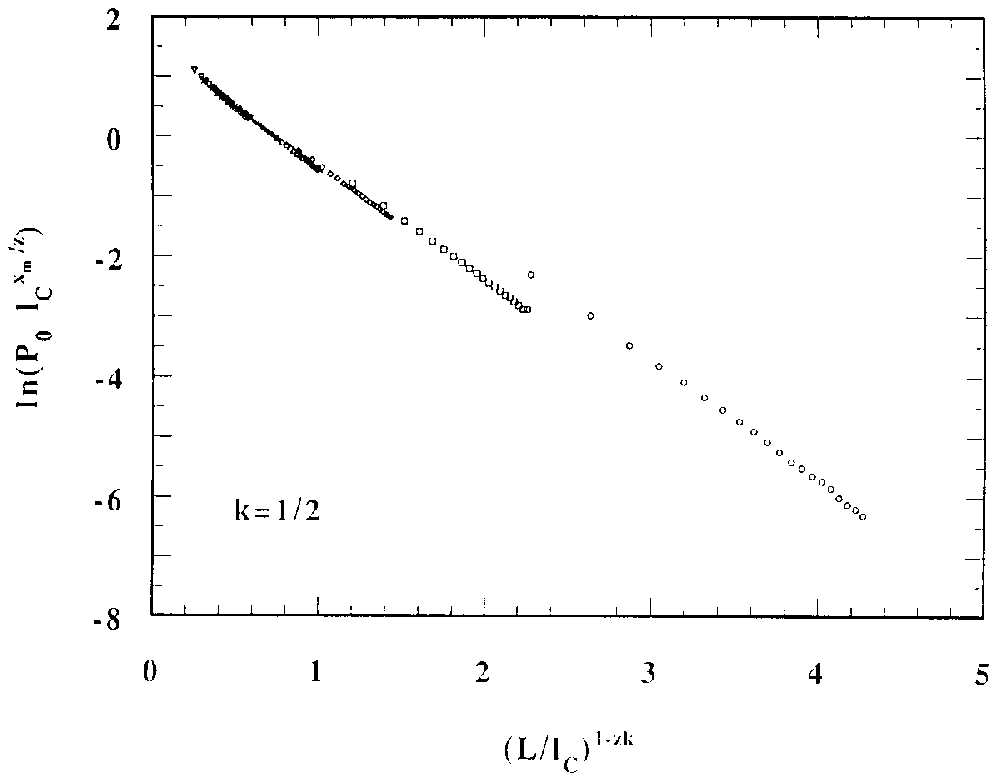}}
\smallskip
\figure{Finite--size scaling of the tip order
parameter at the critical point for a relevant surface shape ($k\!=\!1/2$, same
symbols as in figure 4 for $C$). The almost linear variation of
$\ln P_0 l_C^{x_m/z}$ as a function of $(L/l_C)^{1\!-\!zk}$ indicates a stretched
exponential behaviour of the tip percolation probability.} 
\endinsert 
\endgroup 
\par}

In order to check the expected stretched exponential
behaviour in the relevant case, we performed a
finite--size scaling study at the critical point. The tip percolation
probability, for~$k\!=\!1/2$ and $1\!\leq\! C\!\leq\!4$, was calculated at the
percolation threshold on systems with size $50\!\leq\! L\!\leq\!1000$, taking
averages over $10^5$ MC samples. According to equation~(2.4), $P_0l_C^{x_m/z}$
is a function of~$L/l_C$ which depends only  on~$k$. The scaling behaviour is
shown in figure~7 where the deviations at small $C$--values is likely to be due
to corrections to scaling. Actually, the reference fixed point for which
equation~(2.4) was written corresponds to $C\!\to\!\infty$. In this
semi--logarithmic plot, a linear dependence on $(L/l_C)^{1-zk}$ is obtained. It
corresponds to a stretched exponential behaviour for the scaling function
$m_C(L/l_C)$ although some power in front of the exponential cannot be excluded.
Other simple combinations of~$z$ and~$k$ for the stretching exponent did not
lead to a linear behaviour. 

{\par\begingroup\medskip
\epsfxsize=9truecm
\topinsert
\centerline{\epsfbox{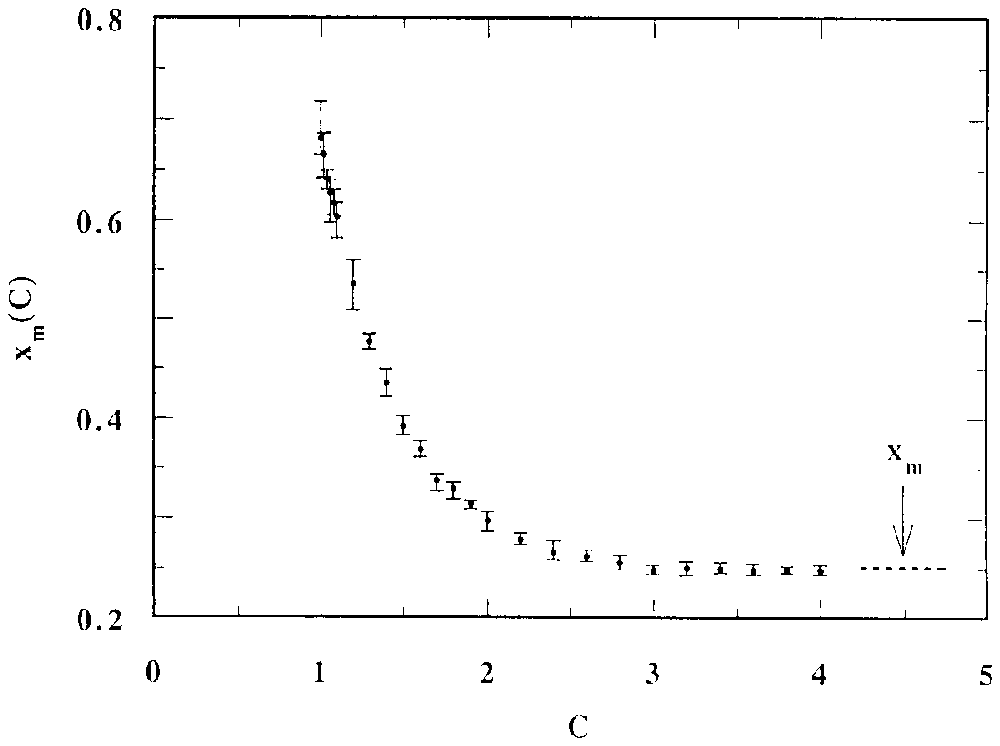}}
\smallskip
\figure{Scaling dimension of the tip order parameter
as a function of $C$ obtained through finite--size scaling when the
shape is marginal ($k\!=\!1/z$). The local exponent $x_m(C)$ diverges when
$C\!\to\!0$ and goes to its unperturbed value $x_m\!=\!0.252(3)$ when
$C\!\to\!\infty$.}
\endinsert  
\endgroup 
\par}

In the marginal case, $k\!=\!1/z$, the scaling dimension of the tip order
parameter is expected to vary with $C$. Equation~(2.5), with $b\!=L^{1/z}$, leads
to the finite--size behaviour $P_0(L)\!\sim\! L^{-x_m(C)/z}$ at $p\!=\! p_c$
which can be used to determine $x_m(C)$. This has been done taking
averages for $P_0(L)$ over $2.10^5$ samples with $L\!=\!500,520,\dots,1000$.
The tip exponent was deduced from the asymptotic slopes of log--log plots. The
results are shown in figure~8. The dependence on $C$ is qualitatively similar to
the mean--field one in figure~2: the scaling dimension diverges when $C$
vanishes and it decreases to the bulk value $x_m\!=\!\beta/\nu\!=\!0.252(3)$
when $C$ goes to infinity.

The fractal structure of finite critical clusters was also studied. Since 
details  can be found in reference~[22], here we only give a brief summary
of our results. As in figure~3, $2.10^5$ percolation clusters starting from the
tip  were generated at the percolation threshold, inside a parabolic system with
size $L\!=\!1000$ for $C\!=\!2,3,4$ and values of $k$ ranging from $0.25$ to
$0.75$. The mean square radii of gyration, $\langle X_s^2\rangle$ in the
$x$--direction and $\langle Y_s^2\rangle$ in the transverse direction, were
determined for $s$--site "finite" clusters, i. e. clusters with $x_{max}\!<\!
L$. 

For an unconfined system, the cluster size behaves as
$$
s\sim\overline{X_s}^{d_\parallel}\sim\overline{Y_s}^{d_\perp}
\eqno(4.2)
$$
where $\overline{X_s}\!=\!\langle X_s^2\rangle^{1/2}$,
$\overline{Y_s}\!=\!\langle Y_s^2\rangle^{1/2}$.
The exponents are fractal dimensions which, extending an argument of
Stauffer~[23], are related to the directed percolation exponents through
$$
{d_\parallel}={\beta+\gamma\over
\nu_\parallel}= 1.473(2)\qquad{d_\perp}={\beta+\gamma\over\nu_\perp}=2.329(3).
\eqno(4.3)
$$
The second value is greater than the dimension of the system due to the
anisotropy as explained in reference~[22]. A single fractal dimension $d_f$ may
be also defined using the characteristic length $l\sim({l_\parallel}
{l_\perp})^{1/2}\sim s^{1/d_f}$ associated with the surface of the cluster so
that~[20] $2/d_f\!=\!1/d_\parallel+1/d_\perp$ and then, $d_f\!=\!1.805(2)\!<\!2$.

{\par\begingroup\medskip
\epsfxsize=9truecm
\topinsert
\centerline{\epsfbox{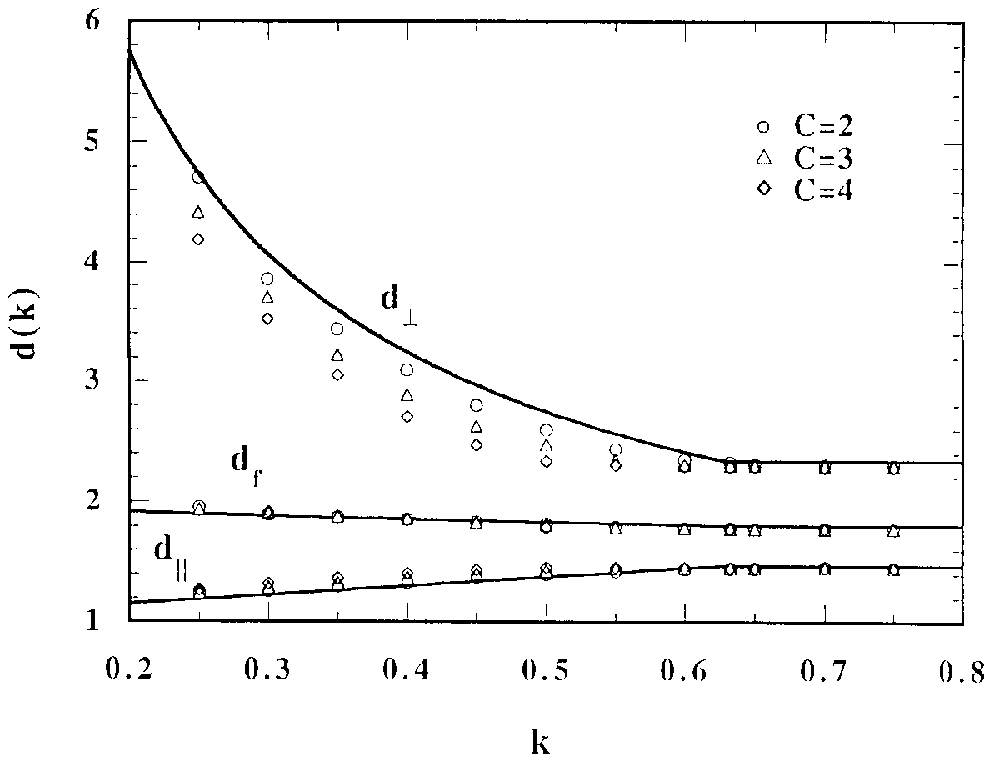}}
\smallskip
\figure{Fractal dimensions versus $k$
for finite critical percolation clusters starting at the tip. The variation in
the relevant regime, $k\!<\!1/z\!=\!0.633(2)$, is well described by the
blob picture approach (lines) when $C$ decreases. The deviations
for large $C$--values are due to the finite size of the system:
only a few blobs can develop and the fractal dimensions are then closer to the
unperturbed ones.}
\endinsert  
\endgroup
\par}

Inside a parabolic system, one expects a similar behaviour with $k$--dependent
fractal dimensions $d_\parallel(k)$, $d_\perp(k)$, when the shape is a relevant
perturbation. Log--log plots of the MC results are linear for cluster sizes
between $2^6$ and $2^{12}$--$2^{13}$. Deviations at larger sizes are due to
finite--size effects. The fractal dimensions are shown in figure~9 as
functions of $k$. The form of the variation in the relevant regime, $k\!<\!1/z$,
can be obtained using a blob picture approach~[24--26,14]. The cluster
configuration is supposed to follow from the piling up of anisotropic blobs
inside the parabolic system. Within each blob the structure is the same as for
unconfined clusters with the unperturbed values of the fractal dimensions. In
this way one obtains~[22]
$$
d_\parallel(k)=1+zk(d_\parallel-1),\qquad d_\perp(k)={d_\parallel(k)\over k},
\qquad 0<k\leq1/z
\eqno(4.4)
$$
in good agreement with the Monte--Carlo data as shown in figure~9. 

\section{Conclusion}

The local critical behaviour for directed percolation has been investigated at
the tip of two--dimensional parabolic--like systems. The problem has been
treated using mean--field theory and MC--simulations. 
In the marginal case, $k\!=\!1/z$, the percolation probability and the order
parameter correlation function display a non--universal local critical
behaviour:  the local exponents vary continuously with the shape of the system.
In the relevant case, $k\!<\!1/z$, the critical tip percolation probability is a
stretched exponential function of the reduced size $L/l_C$ and
finite critical clusters have $k$--dependent fractal dimensions.

Although we were unable, due to strong finite--size effects, to check a
similar stretched exponential dependence on $\xi_\parallel/l_C$ in the
relevant case, the following general picture seems to emerge from our mean--field
and MC results, exact and MC results for the Ising model~[10--13] and ordinary
percolation~[15]: 
\item{$\bullet$} In isotropic critical systems,  conformal transformations,
from the half--space to the parabola, of the correlation function and the order
parameter profile (with appropriate fixed boundary conditions) show that these
quantities become stretched exponential functions of $x/l_C$ or $L/l_C$ with a 
stretching exponent~$1\!-\!k$. According to our mean--field and MC--results, this
exponent should be changed into $1\!-\! zk$ in anisotropic critical systems.
Conformal methods can no longer be used with anisotropic critical systems for
which the conformal group is replaced by the Schr\"odinger group (when
$z\!=\!2$)~[27].  
\item{$\bullet$} In isotropic off--critical systems, the tip order
parameter is a stretched exponential function of $\xi/l_C$
(with there $l_C\!=\!C^{1/(1\!-\! k)}$) with a stretching exponent~$(1\!-\!
k)/k$. This follows from exact results on the Ising model~[11,12] and also from
an heuristic argument assuming that the tip order, which is induced by the bulk
order at a distance~$D\!\sim\!(\xi/C)^{1/k}$ (where the width of the system is
of the order of the bulk correlation length), decays with~$D$ like the critical
tip--to--bulk correlation function~[10,11]. The same argument, applied to the
anisotropic off--critical system, gives a stretched exponential function of
$\xi_\parallel/l_C$ with a stretching exponent $(1\!-\! zk)/zk$. Such a
behaviour could not be checked numerically here, since our finite off--critical
system involves some unknown combination of the two scaling variables $L/l_C$
and $\xi_\parallel/l_c$.  

Further numerical and analytical work on anisotropic systems would be useful
to confirm the last point. 

\ack
We thank S. Blawid and I. Peschel for discussions and collaboration, B. Berche
and J. M. Debierre for communicating their Monte--Carlo algorithm. CK
thanks Henri Poincar\'e University for hospitality. This work was supported by
CNIMAT under project No 155C93.  

\references
\numrefjl{[1]}{Cardy J L 1983}{\JPA}{16}{3617} 
\numrefjl{[2]}{Barber M N, Peschel I and Pearce P A
1984}{J. Stat. Phys.}{37}{497}
\numrefjl{[3]}{Cardy J L 1984}{Nucl. Phys.\ \rm B}{240}{514}
\numrefjl{[4]}{Peschel I 1985}{Phys. Lett.}{110A}{313}
\numrefjl{[5]}{Kaiser C and Peschel I 1989}{J. Stat. Phys.}{54}{567}
\numrefjl{[6]}{Davies B and Peschel I 1991}{\JPA}{24}{1293}
\numrefjl{[7]}{Guttmann A J and Torrie G M 1984}{\JPA}{17}{3539}
\numrefjl{[8]}{Cardy J L and Redner S 1984}{\JPA}{17}{L933}
\numrefjl{[9]}{Duplantier B and Saleur H 1986}{\PRL}{57}{3179}
\numrefjl{[10]}{Igl\'oi F, Peschel I and Turban L 1993}{Adv. Phys.}{42}{683}
\numrefjl{[11]}{Peschel I, Turban L and Igl\'oi F 1991}{\JPA}{24}{L1229}
\numrefjl{[12]}{Davies B and Peschel I 1992}{Ann. Phys. (Leipzig)}{2}{79}
\numrefjl{[13]}{Blawid S and Peschel I 1994}{\ZP\ \rm B}{95}{73}
\numrefjl{[14]}{Turban L and Berche B 1993}{\JP\  I}{3}{925}
\numrefjl{[15]}{Berche B, Debierre J M and Eckle P}{1994}{\PR\ \rm E}{\it (to
appear)}   
\numrefbk{[16]}{Privman V and \v{S}vraki\'c N M 1989}{Directed Models 
of Polymers, Interfaces and Clusters}{Lecture Notes in Physics 
338 (Berlin: Springer)}
\numrefjl{[17]}{Binder K and Wang J S 1989}{J. Stat. Phys.}{55}{87}
\numrefjl{[18]}{Turban L 1992}{\JPA}{25}{L127} 
\numrefjl{[19]}{Igl\'oi F 1992}{\PR\ \rm B}{45}{7024} 
\numrefjl{[20]}{Kinzel W 1983}{Ann. Israel Phys. Soc.}{5}{425}
\numrefjl{[21]}{Essam J W, Guttmann A J and De'Bell K 1988}{\JPA}{21}{3815}
\numrefjl{[22]}{Kaiser C and Turban L 1994}{\JPA}{27}{L579}
\numrefbk{[23]}{Stauffer D 1985}{Introduction to Percolation
Theory}{(London: Taylor \& Francis) p 64}
\numrefjl{[24]}{Pincus P 1976}{Macromolecules}{9}{386}
\numrefbk{[25]}{de Gennes P G 1979}{Scaling Concepts in Polymer
Physics}{(Ithaca: Cornell University Press) p~50}
\numrefjl{[26]}{Turban L and Debierre JM 1984}{\JPA}{17}{L289}
\numrefjl{[27]}{Henkel M 1994}{J. Stat. Phys.}{75}{1023}

\vfill\eject\bye